\documentclass[10pt,conference]{IEEEtran}
\IEEEoverridecommandlockouts
\usepackage[normalem]{ulem}
\usepackage{paralist}
\usepackage{tabularx}
\usepackage{booktabs}
\usepackage{comment}
\usepackage{blindtext}
\usepackage[colorinlistoftodos,prependcaption,textsize=tiny]{todonotes}
\usepackage{regexpatch}
\usepackage{hyperref}
\usepackage{enumitem}
\usepackage[font=footnotesize,labelfont=bf]{caption}
\usepackage{array,multirow,graphicx}
\usepackage{lipsum}
\usepackage{mathtools}
\usepackage{xcolor}
\usepackage[ruled]{algorithm2e}
\usepackage{dirtree}
\usepackage{algpseudocode}
\usepackage{amsmath,amssymb,amsfonts}
\usepackage{graphicx}
\usepackage{cleveref}
\creflabelformat{equation}{#2#1#3}
\usepackage{dblfloatfix}
\usepackage{nicefrac}
\usepackage[caption=false]{subfig}
\usepackage[draft,inline,nomargin,index]{fixme}
\usepackage{textcomp}
\usepackage{mdframed}
\usepackage{tikz}
\usepackage{amssymb}
\usepackage{pifont}
\usepackage{float}
\usepackage{blindtext}

\crefname{lstlisting}{listing}{listings}
\Crefname{lstlisting}{Listing}{Listings}

\usepackage[switch]{lineno}

\newcommand{\xmark}{\ding{55}}%

\newcommand{\hide}[1]{}

\newcommand{\Melissa}{Melissa-DA}
\newcommand{\FTI}{FTI}

\newcommand*\mean[1]{\bar{#1}}

\newtoggle{releaseStuffAfterDoubleBlind}
\toggletrue{releaseStuffAfterDoubleBlind}

\def\BibTeX{{\rm B\kern-.05em{\sc i\kern-.025em b}\kern-.08em
    T\kern-.1667em\lower.7ex\hbox{E}\kern-.125emX}}

\usepackage{listings}
\definecolor{aliceblue}{rgb}{0.94, 0.97, 1.0}
\definecolor{bleudefrance}{rgb}{1.0, 0.44, 0.37}
\definecolor{backcolour}{rgb}{0.8,0.8,0.8}
\definecolor{beige}{rgb}{0.96, 0.96, 0.86}
\usepackage{color}
\definecolor{mygreen}{rgb}{0,0.6,0}
\definecolor{mygray}{rgb}{0.5,0.5,0.5}
\definecolor{mymauve}{rgb}{0.58,0,0.82}
\definecolor{anti-flashwhite}{rgb}{0.95, 0.95, 0.96}

\fxsetup{theme=color,mode=multiuser}
\FXRegisterAuthor{kk}{akk}{\color{orange}\bf Kai}
\FXRegisterAuthor{lb}{alb}{\color{blue}\bf Leo}
\FXRegisterAuthor{mw}{amw}{\color{green}\bf Wahib}

\colorlet{punct}{red!60!black}
\definecolor{background}{HTML}{EEEEEE}
\definecolor{delim}{RGB}{20,105,176}
\colorlet{numb}{magenta!60!black}

\lstdefinelanguage{json}{
    basicstyle=\scriptsize\ttfamily,
    numbers=none,
    showstringspaces=false,
    breaklines=true,
    frame=lines,
    backgroundcolor=\color{anti-flashwhite},
    literate=
     *{0}{{{\color{numb}0}}}{1}
      {1}{{{\color{numb}1}}}{1}
      {2}{{{\color{numb}2}}}{1}
      {3}{{{\color{numb}3}}}{1}
      {4}{{{\color{numb}4}}}{1}
      {5}{{{\color{numb}5}}}{1}
      {6}{{{\color{numb}6}}}{1}
      {7}{{{\color{numb}7}}}{1}
      {8}{{{\color{numb}8}}}{1}
      {9}{{{\color{numb}9}}}{1}
      {:}{{{\color{punct}{:}}}}{1}
      {,}{{{\color{punct}{,}}}}{1}
      {\{}{{{\color{delim}{\{}}}}{1}
      {\}}{{{\color{delim}{\}}}}}{1}
      {[}{{{\color{delim}{[}}}}{1}
      {]}{{{\color{delim}{]}}}}{1},
}

\begin{document}

\title{A Framework for Automatic Validation and Application of Lossy Data Compression in Ensemble Data Assimilation}
\iftoggle{releaseStuffAfterDoubleBlind}{%
\author{\IEEEauthorblockN{Kai Keller\IEEEauthorrefmark{1},
Hisashi Yashiro\IEEEauthorrefmark{2},
Mohamed Wahib\IEEEauthorrefmark{3},
Balazs Gerofi\IEEEauthorrefmark{4},
Adrian Cristal Kestelman\IEEEauthorrefmark{1},
Leonardo Bautista-Gomez\IEEEauthorrefmark{1}}
\IEEEauthorblockA{\IEEEauthorrefmark{1}Barcelona Supercomputing Center (BSC-CNS), Spain
\\\{kai.keller, leonardo.bautista, adrian.cristal\}@bsc.es}
\IEEEauthorblockA{\IEEEauthorrefmark{2}National Institute for Environmental Studies, Japan
\\yashiro.hisashi@nies.go.jp}
\IEEEauthorblockA{\IEEEauthorrefmark{3}National Institute of Advanced Industrial Science and Technology (AIST), Tokyo, Japan
\\mohamed.attia@aist.go.jp}
\IEEEauthorblockA{\IEEEauthorrefmark{4}RIKEN, Japan
\\bgerofi@riken.jp}}
}

\maketitle

\thispagestyle{empty}

\thispagestyle{plain}
\pagestyle{plain}
\begin{abstract}
    Ensemble data assimilation techniques form an indispensable part
    of numerical weather prediction. As the ensemble size grows and
    model resolution increases, the amount of required storage
    becomes a major issue. Data compression schemes may come to the
    rescue not only for operational weather prediction, but also for
    weather history archives. In this paper, we present the design
    and implementation of an easy-to-use framework for evaluating the
    impact of lossy data compression in large scale ensemble data
    assimilation. The framework leverages robust statistical
    qualifiers to determine which compression parameters can be
    safely applied to the climate variables. Furthermore, our
    proposal can be used to apply the best parameters during
    operation, while monitoring data integrity. We perform an
    exemplary study on the Lorenz96 model to identify viable
    compression parameters and achieve a 1/3 saving in storage space
    and an effective speedup of 6\% per assimilation cycle, while
    monitoring the state integrity.
\end{abstract}

\section{Introduction}\label{sec:introduction}

Ensemble methods have become increasingly important for numerical
weather and climate prediction. One of the main reasons for this is
the encoded statistics in the ensemble. Whereas the widely used
four-dimensional variational data assimilation (4D-var) often leads
to more accurate predictions, it does not provide a simple way for
building assumptions on the prediction
uncertainty~\cite{ngodock_ensemble_2020}. Therefore, it is frequently
combined with an ensemble method. On the other hand, ensemble methods
are often preferred from the beginning. For instance, due to the
typically less implementation effort or because they constitute the
better fit in certain cases~\cite{gopalakrishnan_comparison_2019,
    lorenc2003relative,lorenc_potential_2003}.
Ensemble data assimilation with large ensembles and large models
requires high performance I/O~\cite{klower_compressing_2021,
    schnase_big_2016,eggleton_open_2020}. This is due to the large
amount of data that needs to be circulated between different
constituents of the assimilation system. 

A widely used workflow in
ensemble data assimilation is to perform the climate simulation and
the data assimilation on separate executables~\cite{zheng_offline_2020}.
The ensemble members
(i.e., the climate simulations) store the climate states to the file
system, and after all simulations have finished, the data
assimilation system reads the files, assimilates the observations and
writes back the improved states to storage. The ensemble members
then reread them to perform the next assimilation cycle. \hide{Some
frameworks avoid the state circulation through the IO layer and use
the network instead to transfer the states between the simulations
and assimilation system. In any case, } The amount of data
transferred between the two steps often leads to an I/O bottleneck, where
the storage subsystem cannot deliver the throughput that is needed to keep up
with the computing power. In some cases, I/O can be overlapped with computation and be performed
in the background, which alleviates the I/O overhead itself. The
states can also be transferred through the network, bypassing the I/O
layer. However, this approach is limited by the memory available and
raises fault tolerance issues, since such approaches typically require
large monolithic MPI allocations.

However, reducing the data to minimize storage requirements and storage space
availability for other users is still beneficial. A recent example of
storage based state circulation between simulation and data
assimilation system has been published by Yashiro et al
.~\cite{terasaki_big}. The article presents the execution of the
NICAM-LETKF system on Fugaku~\cite{okazaki_supercomputer_2020}, using 82\% of the entire system. During
each cycle, the system circulates more than 400~TB of data through
the parallel file system. Yashiro et al. also compare double to mixed
precision executions, where the computing times with mixed precision show a 1.6x speedup
compared to double precision. This demonstrates the prospect of
departing from the double precision doctrine in climate science.
There are a number of works studying explicitly the impact and
advantage of mixed precision in data
assimilation~\cite{hatfield_improving_2018,hatfield_choosing_2018,nakano_single_2018}

Besides using mixed precision, data can be reduced by using compression schemes. Data can be compressed in a lossless fashion, i.e., without any loss in information, or through lossy methods, i.e.,
with a certain loss in accuracy. Since climate systems are highly
non-linear and typically 
chaotic, it is essential not to introduce
significant perturbations when applying compression to the states. 
On the other hand, the models and the observations are both imperfect
and inevitably introduce errors to the states. This means that the
intrinsic precision of the states can be lower than the precision of
the data type used in the application. Indeed, multiple previous
studies have shown that climate simulations can tolerate certain loss
in data precision~\cite{abdulah_accelerating_2022,
    baker_methodology_2014,baker_evaluating_2016,
    poppick_statistical_2020}.

In this work, we propose a framework that:
\begin{inparaenum}[(1)]
    \item explores the impact of lossy data compression of the
    climate states and the ensemble consistency through time (i.e., the error propagation), and
    \item dynamically selects the best compression parameters during operational mode.
\end{inparaenum}
The former constitutes the \emph{validation mode} of the framework and
the latter is the \emph{dynamic mode}. In the validation mode, we generate
and store validation data each cycle. 
The collected data provides a means to evaluate the impact of inconsistency
in the states and ensemble through time.  We provide a JSON
configuration file to conveniently set the desired compression parameters. During the
dynamic mode, the compression parameters are applied to the states to
reduce their size before writing them to the file system. Optionally,
the states can be tested by a validation function, before
writing them, to ensure their consistency. The framework allows easy
interfacing and is build on top of the ensemble data assimilation
architecture of~\Melissa{}. Thus, after exposing the
simulation variables to~\Melissa{} and a few more steps,
it can be operated with the two
modes from above, in combination with the wide range of ensemble data
assimilation methods that are contained in~\Melissa{}.

The rest of the paper is organized as follows.
Section~\ref{sec:background} provides background information on ensemble data assimilation techniques.
Section~\ref{sec:Implementation} presents the design and
implementation of the proposed framework and explains its usage, and
Section~\ref{sec:evaluation} provides experimental evaluation.
Related work is surveyed in Section~\ref{sec:relatedwork}.
Finally, Section~\ref{sec:conclusion} concludes the paper.

\section{Background}\label{sec:background}
In this section, we outline the basic concepts behind our work
and clarify the terminology that we use.

\subsection{Ensemble Data Assimilation}\label{subsec:ensemble-data-assimilation}

Data assimilation is based on Bayes' theorem, allowing us to combine the
information from both real world observation and numerical model states.
Combining the information from both sources leads to an improved accuracy of the state~\cite{wikle_bayesian_2007}.
Ensemble data assimilation follows a Monte Carlo approach,
approximating the state mean and covariance by the moments of
a statistically significant sample of states.
The most common ensemble methods for data assimilation are the
ensemble Kalman filter (EnKF)~\cite{evensen_sequential_1994} and
the particle filter (PF) ~\cite{van_leeuwen_particle_2019}. To address
the issues that arise from relatively small ensemble sizes given the
high dimensionality of the climate states, modified flavors of the
original versions are used, e.g., LETKF~\cite{hunt_efficient_2007} or LAPF~\cite{potthast_localized_2019}, and others.

One important difference between particle filtering and ensemble Kalman filtering is that the particle states do not change during the filter update. 
The particles get assigned a weight and particles with small
weights are discarded. Particles with high weights will be selected for
propagation during the next cycle. The particle filter implementation
in~\Melissa{} uses \emph{sequential importance resampling} (SIR), where
particles with high weights are multiplied to avoid ensemble
shrinking. This technique relies on a randomization of the model,
reaching different output states when starting from identical
particles. This is achieved either by randomization during the model
evolution, or by introducing small perturbations to the input particle
states before the propagation~\cite{leeuwen_particle_2009,van_leeuwen_particle_2019}.

\subsection{Terminology}\label{subsec:terminology}

In spite of the same origin, being both Monte Carlo methods, the
terminology used for ensemble Kalman filters and particle filters is
quite different. To avoid confusion, we will introduce the
terminology that we use in this work. We implemented our validation
framework into the particle filter of \Melissa{}~\footnote{The
validation methods that we use, however, are perfectly suitable for other
ensemble methods as well.}. For this reason, we will use the particle
filter terminology. We refer to a simulation state as particle state,
or sometimes just particle or state. The workflow of ensemble data
assimilation is divided into assimilation cycles. Each cycle
comprises the propagation step and the update step. During the
propagation step, the climate states are advanced by the numerical
model (a.k.a., \emph{propagated}). During the update step, the model states are improved by the
filter update, assimilating the observations. In particle filtering,
the update step is called sampling, or resampling.

\section{Design and Implementation}\label{sec:Implementation}
In this section, we first introduce the particle filter implementation of \Melissa{},
which serves as testbed for our work, and then explain
implementation and operation of our proposed framework in detail.

\subsection{\Melissa{} Particle Filter}\label{subsec:melissa-particle-filter}

\Melissa{} is developed to perform ensemble data assimilation at large
scale. The framework comprises three modules; a launcher, a server, and
multiple runners. In comparison to the common practice in ensemble data
assimilation leveraging a bash script for the workflow, the launcher
replaces the script and orchestrates the submission of the ensemble
simulations and the data assimilation system. The launcher has plugins
for the most common cluster schedulers to submit and monitor the jobs
for the server and runners. In this architecture, the runners constitute a worker
pool for propagating the ensemble states. The server requests available
runners to perform propagation of unscheduled states and performs the
data assimilation step. To ensure optimal fault tolerance, the server
and each runner is allocated on separate jobs and are restarted by the launcher upon failures.
\Melissa{} provides data assimilation with various flavors of ensemble
Kalman filters and allows the creation of custom filter plugins.
Besides the Kalman filter, the framework implements a particle filter
that uses a fast distributed cache on the runner nodes, where particles can
be asynchronously prefetched and cached for future
propagations. The cache is maintained by dedicated MPI processes, asynchronously, and uses
one process per runner node. Generally, for fault tolerance, the states
need to be stored on global storage. Since this is typically slower than
using the local storage, the simulation processes only write and load
from local storage. The cache controller is responsible for
providing the states locally and sending the states to global storage
after they have been propagated.

In most particle filters, in contrast to Kalman
filters, the states are not changed during the filter step. Instead,
states that carry high weights (i.e., that are consistent with the
observation data), are selected for the next assimilation cycle and
states with low weights are discarded. This is precisely why the local
cache helps to overcome the I/O bottleneck; states that have been
selected for the next assimilation are
still locally available on runners that have
propagated them. Since the states remain unchanged during
filtering, they are immediately ready for propagation.
The local cache leverages \FTI{} to store and load
the states. \FTI{} is a multi-level checkpoint/restart library that is
aware of the node local storage. We implemented several compression
techniques into \FTI{} to enable the state compression while
storing the state and decompression while loading it.

\hide{
In the following
sections, we will describe in detail, what our proposed framework provides and
how to configure the different operational modes and compression
parameters.
}

\subsection{High-Level View on the Validation Framework}
\label{subsec:high-level-view-on-the-extensions}

Our proposed framework comprises two modes of operation. The
first mode allows us to explore the impact of data compression on the
integrity of the ensemble and of the states in particular. The second mode allows us to apply the best compression
parameters during operation to increase performance, while respecting
data consistency. We refer to the first mode as \emph{validation mode} and
to the second as \emph{dynamic mode}. The respective modes and
corresponding parameters are selected by providing a JSON configuration
file (see~\Cref{lst:jsoncfg}).

\begin{lstlisting}[language=json,firstnumber=1,
	caption={
  Example of a configuration file. We set the compression parameters for
  two variables named \emph{state1} and \emph{state2}.  The framework
  will operate in validation mode, since the method key is set to
  \emph{validation}. To apply the dynamic mode, the method must be set to
  \emph{dynamic}.
	},
	captionpos=b,label={lst:jsoncfg}]
{
  "variables" : [ "state1", "state2" ],
  "compression" : {
    "method" : "validation",
    "validation" : [
      {
        "mode" : "fpzip",
        "parameter" : [16,24,32]
      },
      {
        "mode" : "zfp",
        "type" : "precision",
        "parameter" : [32,40]
      }
    ],
    "dynamic" : [
      {
        "name" : "state1",
        "sigma" : 10e-7,
        "mode" : "zfp",
        "type" : "accuracy",
        "parameter" : [0,6,8,10]
      },
      {
        "name" : "state2",
        "sigma" : 10e-5,
        "mode" : "fpzip",
        "parameter" : [24,28,32,36,40]
      }
    ]
  }
}
\end{lstlisting} 

Our framework operates in conjunction with the particle filter of \Melissa{}.
During the particle filter update, the particle weights are normalized and $P$
particles are drawn from the resulting distribution function.  $P$
remains constant during all cycles and it can be set in the \Melissa{}
configuration. Hence, during each cycle we propagate the same number of particles.
However, the SIR algorithm leads to a sample of only $M$ distinct particles wiuth typically $M < P$.
Therefore, some particles are multiplied. To account for this,
the model needs to provide some randomness to ensure that two
propagations of the same particle lead to distinct output states

During the validation mode of our framework, we now propagate $(C+1)
\cdot{}P$ particles, with $C$ being the number of parameters that are
specified in the configuration file. The configuration shown
in ~\Cref{lst:jsoncfg}, leads to the propagation of 6 ensembles with
$P$ particles each. One ensemble for the original states and 5
ensembles that use data compression with the specified parameters.
This allows us to compare the ensembles with compressed data to the
original ensemble. Furthermore, we can observe the evolution of each
particle ensemble over time for the number of cycles specified in
the \Melissa{} configuration.

In the dynamic mode we aim to improve the performance of the data
assimilation system. Thus, only one ensemble with $P$ particles is
propagated. During the first assimilation cycle, the compression
parameters are tested using a validation function. Per
default, we check the point-wise maximum error between the compressed
and uncompressed state, however, a custom function can be provided by
the user, or the validation can also be deactivated entirely. If
using the default validation,
the compression parameters that lead to a value larger than
\emph{sigma} (see~\Cref{lst:jsoncfg}) are discarded. The selected
parameters are then stored and ordered by the compression rate.
In subsequent cycles, the best compression parameters are
successively checked by the validation function, and the first
parameter that passes is used to compress the state before writing it
to the file system. We allow setting the compression parameters for
each variable independently. The meta-data that is required to recover the variable with the correct compression settings
is maintained by \FTI{}. A speedup is achieved, if the
compression/decompression plus validation
time is less than the time we save due to write/read the fewer (i.e., compressed) data.

\begin{figure}[t]
  \includegraphics[width=\linewidth]{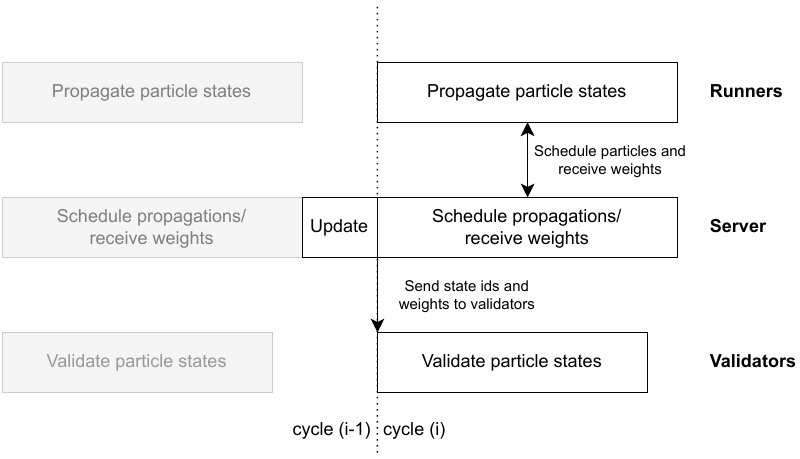}
  \caption{Workflow in validation mode.}\label{fig:validateWorkflow}
\end{figure}

\subsection{Validation Mode}\label{subsec:validate-mode}

For the validation mode, we leverage the \Melissa{} launcher to submit
several validator instances, in addition to the \Melissa{} server and runners. As for the
runners, the number of validators can be specified in the \Melissa{}
configuration. Each validator is executed on one node and is
parallelized leveraging all available cores on the node.~\Cref{fig:validateWorkflow}
shows the workflow in validation mode. The diagram indicates that the propagation of
particles by the runners is overlapped with the validation performed
on the validators. The workflow is a follows, during the propagation
phase, the server schedules particles to the runners and waits for
the particles weights to be returned. As soon as all particles have been
propagated, the server performs the update and communicates particle
ids and weights to the validators. The validators can then start with
the calculation of the statistical qualifiers, where the validation always
processes the states from the previous cycle.

The statistical qualifiers that we calculate are the same as in the work
from Baker et al.~\cite{baker_evaluating_2016}, where the impact
of data compression on climate states is discussed in detail. We
implemented the root mean squared Z-value. Where we calculate both
the value $Z_{x_{c,i}}^{p,0}$, which encodes information on the ensemble spread,
and $Z_{x_{c,i}}^{p,+}$, which can detect a bias on the ensemble
spread introduced by the compression. The two values are calculated
as follows:
\begin{align}
Z_{x_{c,i}}^{p,0}&=\frac{x_{c,i}^p-\mean{x}_{0,i}^{P/p}}{\sigma_{0,
i}^{P/p}}\label{eq:z_value} \\
Z_{x_{c,i}}^{p,+}&=\frac{x_{c,i}^p-\mean{x}_{c,i}^{P/p}}{\sigma_{c,
i}^{P/p}}\label{eq:z_value_bias} \\
\text{RMSZ}_{X_c}^p &= \sqrt{\frac{1}{N}\sum_i^N \left( Z_{x_{c,i}}^p
\right)
^2}\label{eq:rmsz}
\end{align}
with:
\begin{align}
  \mean{x}_{c,i}^{P/p}&=\frac{\sum_{k!=p}^{P}w_k
  x_{c,i}^k}{\sum_{k!=p}^{P}w_k}\\
  \sigma_{c,i}^{P/p}&=\frac{\sum_{k!=p}^{P}w_k\left
  (x_{c,i}^k-\mean{x}_{c,i}^{P/p}\right)^2
  }{\sum_{k!=p}^{P}w_k}
\end{align}
$\text{RMSZ}_{X_c}^p$ is the root mean squared z-value, and
constitutes the quantity that we actually evaluate.
$P$ is the number of particles, $X_c$ specifies the state
variable with $c$ being the compression parameter-id and $c=0$
indicating the uncompressed state. The index $p$ denotes the
particle-id, thus, we have one $\text{RMSZ}_{X_c}^p$ value
for each particle, state variable and compression parameter. Finally $w_p$ denotes the
weight of particle $p$. We further compute the peak signal-to-noise
ratio (PSNR):

\begin{equation}
  \text{PSNR}_{X_c} = 20 \log_{10}\left( \frac{\max \left(
  \left| x_{0,i} \right| \right)
  }{\text{RMSE}_{X_c}}\right)\label{eq:psnr}
\end{equation}
Where $x_{0,i}$ indicate the components of the uncompressed state
variable $X_0$ and $\text{RMSE}_{X_c}$ is the root mean squared error
of the compressed to uncompressed
state variable. Further, we compute the Pearson correlation
coefficient:
\begin{equation}
  \rho_{X_c} = \frac{\sum_i^N\left(x_{0,i}-\mean{x}_{0,i}\right)\left(x_{c,i}-\mean{x}_{c,i}\right)}
  {\sqrt{\sum_i^N\left(x_{0,i}-\mean{x}_{0,i}\right)^2\sum_i^N\left
  (x_{c,i}-\mean{x}_{c,i}\right)^2}}\label{eq:pearson_correlation_coefficient}
\end{equation}
The pointwise maximum error:
\begin{equation}
  \Delta X_c^{\max} = \max\left( \left| x_{0,i} - x_{c,i} \right|
  \right)\label{eq:maximum_pointwise_error}
\end{equation}
And further the mean, standard deviation, minimum and maximum
values for all state variables $X_c$, and the compression rate:
\begin{equation}
  \text{CR}_c = \frac{\text{original size}}{\text{compressed size}}
  \label{eq:compression_rate}
\end{equation}

The validator is implemented in python and can be further customized by
passing custom validation functions to the validator class. The path to
the custom validator script is set in the \Melissa{} configuration.
If no path is set in the configuration, the framework uses a default validator, calculating
the introduced qualifiers.~\Cref{lst:validator-script} shows an
example of a custom validator script.

\begin{figure*}[htb]
  \centering
  \subfloat[FPZIP\label{fig:z_value_fpzip}]{%
    \includegraphics[width=0.475\linewidth]{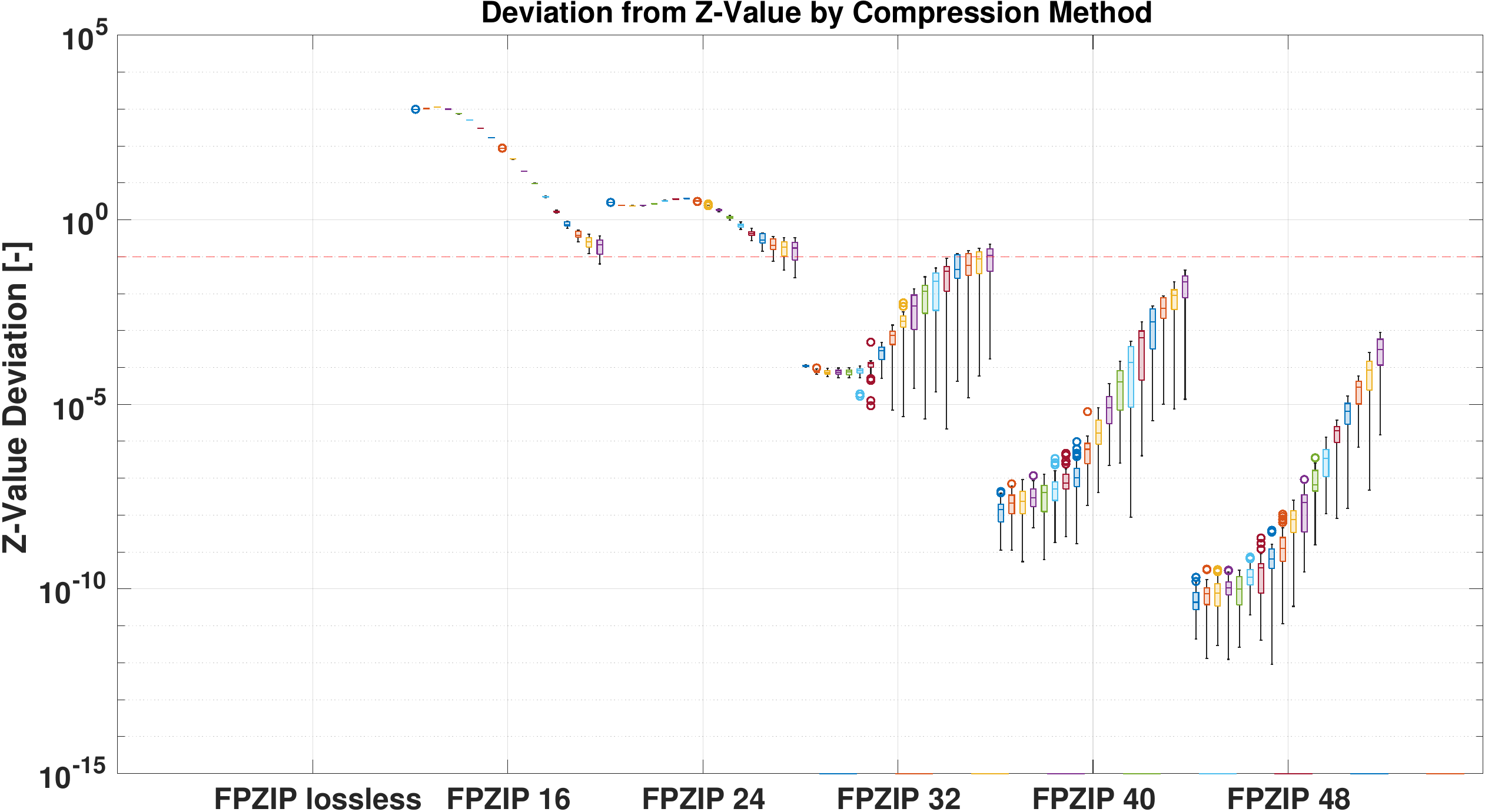}}
  \subfloat[ZFP - Precision\label{fig:z_value_zfp_precision}]{%
    \includegraphics[width=0.475\linewidth]{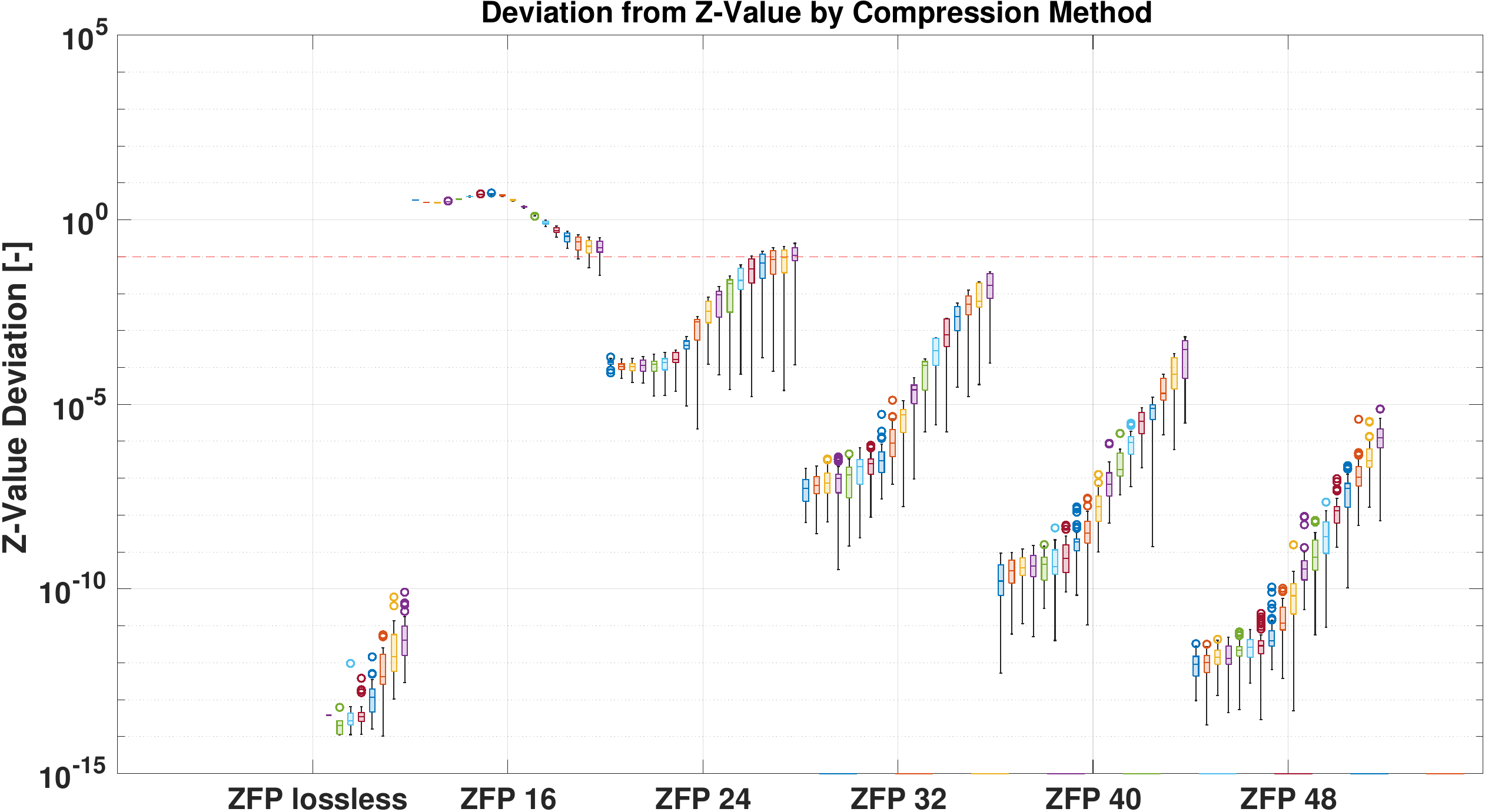}}
  \\
  \subfloat[ZFP - Accuracy\label{fig:z_value_zfp_accuracy}]{%
    \includegraphics[width=0.475\linewidth]{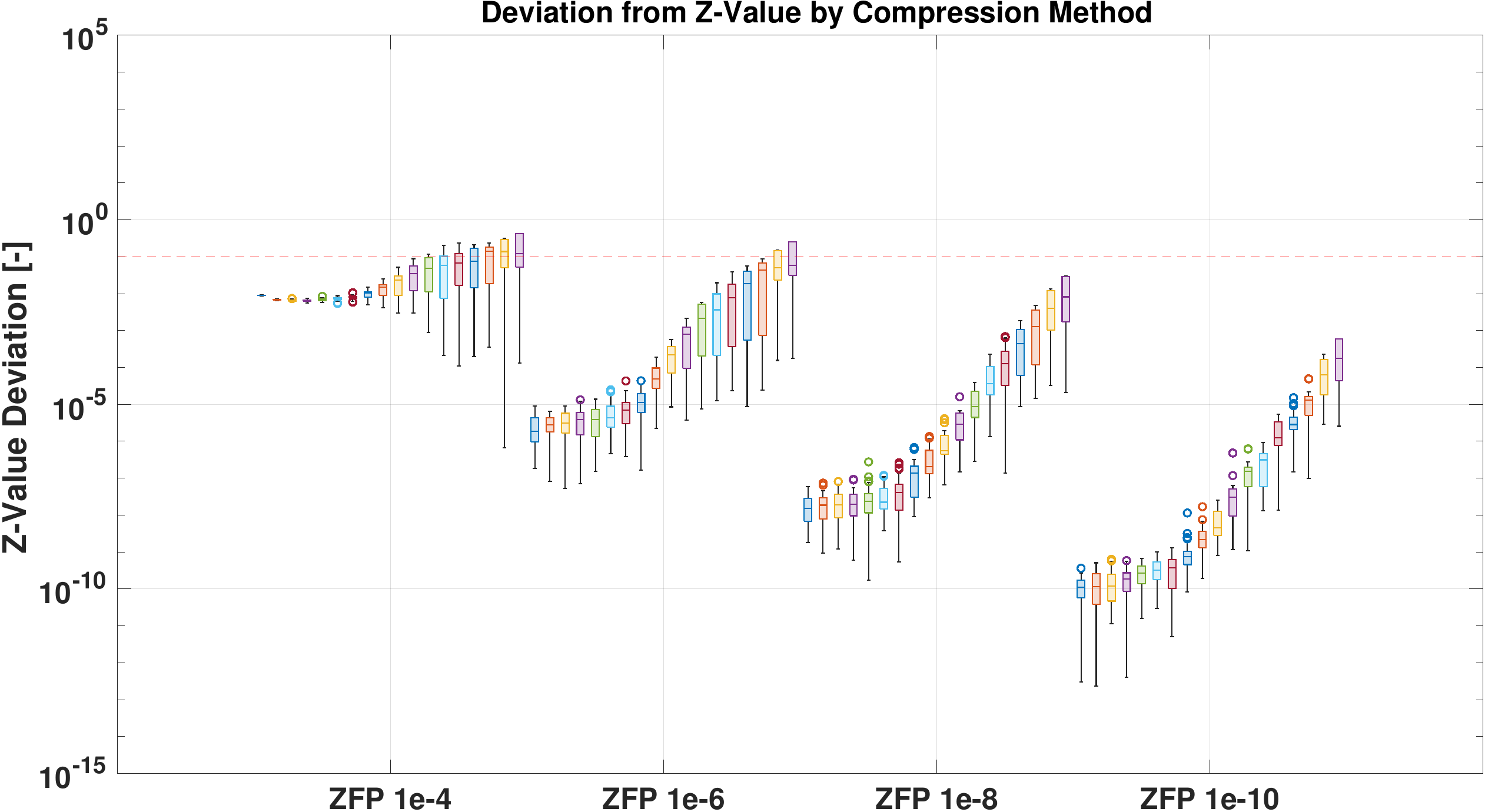}}
  \subfloat[Half/Single precision\label{fig:z_value_hp_sp}]{%
    \includegraphics[width=0.475\linewidth]{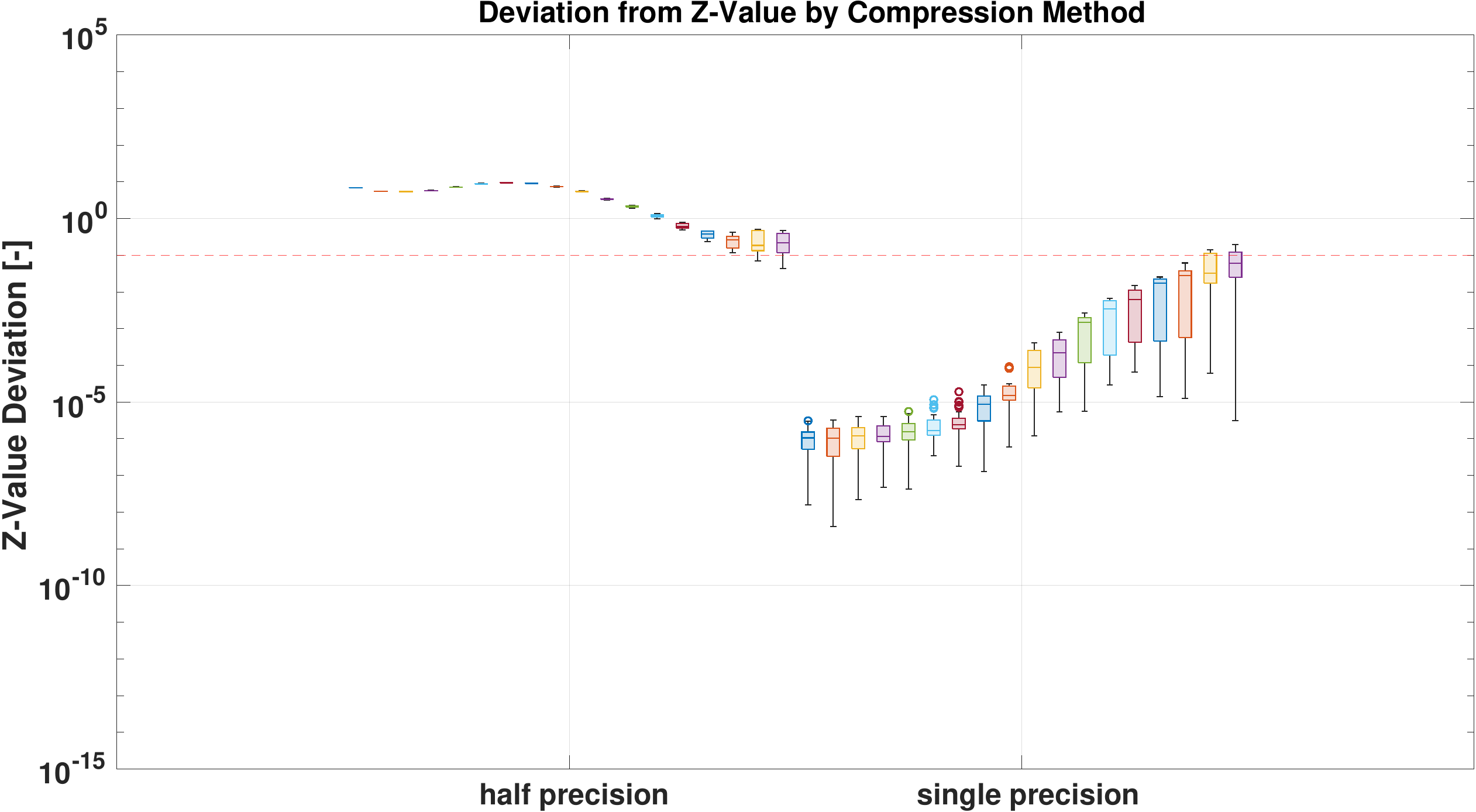}}
  \caption{Z-Value deviation, $\Delta\text{RMSZ}_{X_c}^p$~\ref{eq:z_value_deviation}, for (a) FPZIP, (b) ZFP in precision mode, (c) ZFP in accuracy mode, and (d) half and single precision. The colors indicate values at different cycles.}
  \label{fig:z_value}
\end{figure*}

\begin{lstlisting}[language=python,firstnumber=1,
  caption={
  Example of a custom validator script. The Validator class provides
  arguments for callback functions. The functions have to comply the
  function interface. We can pass the functions as scalar variables
  or arrays. The evaluate and compare functions must return a scalar
  value. The result will be included in the data output. A function
  to write the mean particle state into files can also be provided.
  },
  captionpos=b,label={lst:validator-script}]
from validator import *

def custom_write( mean, variance, cycle, nranks, ndims ):
  '''
  mean      - { "variable1" : [mean_rank1, ...],
                "variable2" : [mean_rank1, ...], ... }
  variance  - { "variable1" : [variance_rank1, ...],
                "variable2" : [variance_rank1, ...], ... }
  cycle     - Assimilation cycle
  nranks    - Number of processes simulation
  ndims     - { "variable1" : [ndim_rank1, ...],
                "variable2" : [ndim_rank1, ...], ... }
  '''
  ...

def custom_evaluate( data, rank, name ):
  '''
  data - variable data on application rank
  rank - application rank
  name - variable name
  '''
  ...

def custom_compare( data, rank, name ):
  '''
  data[0] - uncompressed variable data on application rank
  data[1] - compressed variable data on application rank
  rank    - application rank
  name    - variable name
  '''
  ...

validator = Validator(
  evaluate_function=custom_evaluate,
  compare_function=custom_compare
  write_funtion=custom_write
)
validator.run()
\end{lstlisting}

We provide two kinds of custom functions. The \emph{evaluation
function} is called with the state data, compressed with parameter
$c$ and can be used to compute scalar quantities for
the state variables $X_c$, for instance, the total energy, energy budgets, etc. The
\emph{compare function} is called with both the data of the uncompressed
and the data of the corresponding compressed state and provides a
means to customize the validation, calculating additional qualifiers
for testing the consistency of the compressed states. In addition to
the functions shown in the listing, it is also mandatory to pass
appropriate reduction functions, since the functions are called with
the rank local parts of the data, allowing for computing the quantities in parallel.
All values that are calculated will be recorded and written into a
comma seperated csv file.

\subsection{Dynamic Mode}\label{subsec:adapted-mode}

After identifying the parameters that can safely be applied for data
compression leveraging the validation mode, the
framework can now be operated in dynamic mode. This mode aims the
best performance for the data assimilation using the \Melissa{} particle
filter. The same  configuration file as for the validate mode serves
here as well for providing the compression parameters
(see~\Cref{lst:jsoncfg}). The  dynamic mode operates in two phases.
During the initial phase
we check all compression parameters using a validation function and
discard those parameters that fail the validation. The remaining
parameters are sorted by the compression rate and kept in a variable.
After the initialization, before writing the particle state to disk, we
check the integrity of the state with the validation function after
the compressing it with the best parameter. If the
validation passes, the state is stored to the file system. If
the test fails, we check the next parameter in the list and we repeat
this until we find a parameter that passes the validation. This ensures,
that we never use a compression parameter that introduces
inconsistencies to the particle state.

Per default, the validators are inactive during the dynamic mode.
However, the validators can be activated to compute certain
quantities using custom evaluation functions, or they can be used
to write out the mean particle state with a custom function. This has the
advantage, that those tasks are performed asynchronously to the particle
propagation. The
validators are implemented in Python, hence, we can leverage Python
bindings for common I/O libraries (e.g.,
netCDF4~\cite{rew_unidata_1989}, ADIOS~\cite{noauthor_adios_nodate}
or HDF5~\cite{noauthor_hdf5_nodate})
to write the state data.

\section{Evaluation}\label{sec:evaluation}

\begin{figure*}[htb]
    \centering
    \subfloat[FPZIP, HP, SP\label{fig:npme_nrmse_fpzip_hp_sp}]{%
        \includegraphics[width=0.95\textwidth]{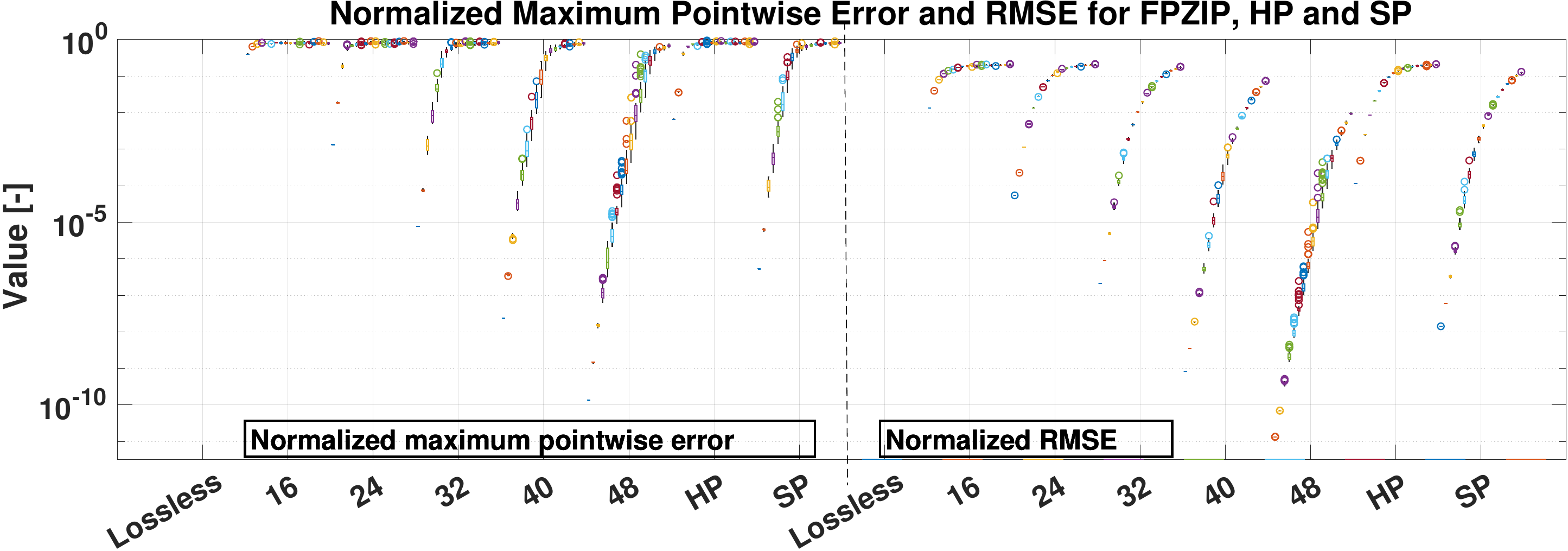}} \\
    \subfloat[ZFP\label{fig:npme_nrmse_zfp}]{%
        \includegraphics[width=0.95\textwidth]{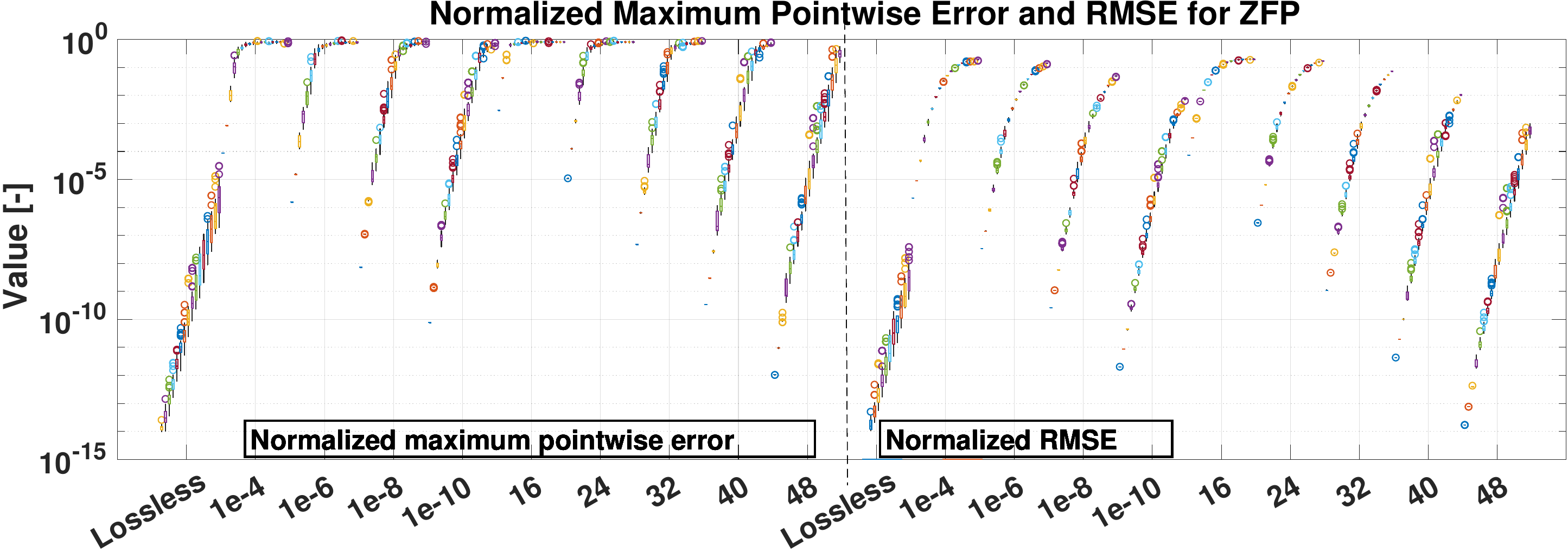}}
    \caption{Normalized maximum pointwise error and normalized root mean square error for (a) FPZIP, half and single precision, and (b) ZFP in accuracy and precision modes. The colors indicate values at different cycles.}
    \label{fig:npme_nrmse}
\end{figure*}

\begin{table*}[htb]
    \centering
    \begin{tabularx}{\linewidth}{cccccccccccc}
        \toprule
        \bf{Cycle} & \bf{FPZIP 32} & \bf{FPZIP 40} & \bf{FPZIP 48} & \bf{ZFP 32} & \bf{ZFP 40} & \bf{ZFP 48} & \bf{ZFP 1e-6} & \bf{ZFP 1e-8} & \bf{ZFP 1e-10} & \bf{HP} & \bf{SP} \\
        \midrule
        \multicolumn{12}{c}{\bf{NRMSE}} \\
        \midrule
        1 & 2.12e-07 & 8.32e-10 & 3.25e-12 & 1.11e-09 & 4.34e-12 & 1.69e-14 & 3.35e-08 & 2.59e-10 & 2.03e-12 & 1.16e-04 & 1.42e-08 \\
        3 & 4.88e-06 & 1.87e-08 & 6.90e-11 & 2.34e-08 & 1.02e-10 & 3.91e-13 & 8.02e-07 & 5.82e-09 & 4.39e-11 & 2.47e-03 & 3.31e-07 \\
        5 & 1.25e-04 & 5.29e-07 & 2.23e-09 & 6.47e-07 & 3.16e-09 & 1.35e-11 & 2.19e-05 & 1.55e-07 & 1.21e-09 & 2.10e-02 & 8.88e-06 \\
        7 & 1.88e-03 & 1.17e-05 & 5.36e-08 & 1.31e-05 & 6.32e-08 & 2.21e-10 & 4.58e-04 & 3.39e-06 & 2.57e-08 & 6.58e-02 & 2.09e-04 \\
        9 & 1.04e-02 & 1.99e-04 & 8.44e-07 & 2.39e-04 & 1.10e-06 & 5.75e-09 & 3.37e-03 & 6.34e-05 & 5.36e-07 & 1.20e-01 & 1.90e-03 \\
        11 & 3.32e-02 & 1.68e-03 & 2.20e-05 & 1.97e-03 & 4.64e-05 & 3.38e-07 & 1.37e-02 & 7.48e-04 & 1.04e-05 & 1.60e-01 & 8.83e-03 \\
        13 & 7.27e-02 & 7.45e-03 & 2.09e-04 & 8.19e-03 & 2.26e-04 & 1.42e-06 & 3.82e-02 & 3.93e-03 & 1.42e-04 & 1.84e-01 & 2.73e-02 \\
        15 & 1.18e-01 & 2.25e-02 & 1.36e-03 & 2.44e-02 & 1.55e-03 & 2.00e-05 & 7.71e-02 & 1.34e-02 & 9.56e-04 & 1.94e-01 & 6.02e-02 \\
        \midrule
        \multicolumn{12}{c}{\bf{NPME}} \\
        \midrule
        1 & 7.68e-06 & 2.34e-08 & 1.33e-10 & 4.74e-08 & 3.44e-10 & 1.05e-12 & 1.54e-06 & 7.20e-09 & 7.62e-11 & 6.56e-03 & 5.26e-07 \\
        3 & 1.38e-03 & 4.38e-06 & 1.52e-08 & 3.81e-06 & 2.66e-08 & 9.60e-11 & 2.52e-04 & 1.88e-06 & 8.73e-09 & 4.11e-01 & 1.02e-04 \\
        5 & 4.86e-02 & 2.34e-04 & 1.11e-06 & 3.01e-04 & 1.68e-06 & 8.27e-09 & 1.06e-02 & 6.61e-05 & 5.37e-07 & 7.35e-01 & 3.80e-03 \\
        7 & 4.77e-01 & 6.16e-03 & 3.15e-05 & 5.90e-03 & 3.42e-05 & 1.10e-07 & 2.25e-01 & 1.77e-03 & 1.24e-05 & 7.72e-01 & 1.17e-01 \\
        9 & 6.76e-01 & 1.04e-01 & 5.05e-04 & 1.31e-01 & 6.60e-04 & 4.09e-06 & 5.64e-01 & 3.68e-02 & 3.10e-04 & 8.00e-01 & 4.97e-01 \\
        11 & 7.47e-01 & 5.05e-01 & 1.73e-02 & 5.19e-01 & 3.01e-02 & 2.35e-04 & 6.70e-01 & 3.59e-01 & 7.01e-03 & 8.24e-01 & 6.47e-01 \\
        13 & 8.00e-01 & 6.35e-01 & 1.27e-01 & 6.59e-01 & 1.31e-01 & 1.04e-03 & 7.53e-01 & 5.80e-01 & 9.36e-02 & 8.08e-01 & 7.38e-01 \\
        15 & 7.96e-01 & 7.24e-01 & 4.59e-01 & 7.15e-01 & 4.57e-01 & 1.29e-02 & 7.80e-01 & 6.83e-01 & 3.73e-01 & 8.05e-01 & 7.79e-01 \\
        \midrule
        \multicolumn{12}{c}{\bf{Pearson Correlation Coefficient}} \\
        \midrule
        1  & 1 & 1 & 1 & 1 & 1 & 1 & 1 & 1 & 1 & 1 & 1 \\
        3  & 1 & 1 & 1 & 1 & 1 & 1 & 1 & 1 & 1 & 0.99986 & 1 \\
        5  & 1 & 1 & 1 & 1 & 1 & 1 & 1 & 1 & 1 & 0.98953 & 1 \\
        7  & 0.99992 & 1 & 1 & 1 & 1 & 1 & 0.99999 & 1 & 1 & 0.89889 & 1 \\
        9  & 0.99750 & 1 & 1 & 1 & 1 & 1 & 0.99974 & 1 & 1 & 0.66111 & 0.99991 \\
        11 & 0.97405 & 0.99993 & 1 & 0.99991 & 1 & 1 & 0.99550 & 0.99999 & 1 & 0.38879 & 0.99815 \\
        13 & 0.87646 & 0.99870 & 1 & 0.99839 & 1 & 1 & 0.96534 & 0.99963 & 1 & 0.20570 & 0.98261 \\
        15 & 0.67014 & 0.98790 & 0.99995 & 0.98595 & 0.99994 & 1 & 0.85887 & 0.99574 & 0.99998 & 0.11110 & 0.91457 \\
        \bottomrule
    \end{tabularx}
    \caption{Average values of the statistical qualifiers NRMSE, NPME and $\rho_{X_c}$ for selected compression parameters. The rows show the evolution of the qualifiers by assimilation cycles.}
    \label{table:statistics}
\end{table*}

\begin{table*}[b]
    \centering
    \begin{tabularx}{\linewidth}{cccccccccccc}
        \toprule
        \multicolumn{12}{c}{\bf{Compression Rate}} \\
        \midrule
        \bf{State Size [MB]} & \bf{FPZIP 32} & \bf{FPZIP 40} & \bf{FPZIP 48} & \bf{ZFP 32} & \bf{ZFP 40} & \bf{ZFP 48} & \bf{ZFP 1e-6} & \bf{ZFP 1e-8} & \bf{ZFP 1e-10} & \bf{HP} & \bf{SP} \\
        \midrule
        16 & 2.332125 & 1.805610 & 1.473040 & 1.897811 & 1.533923 & 1.287127 & 2.240469 & 1.799499 & 1.503566 & 4 & 2 \\
        32 & 2.332534 & 1.805876 & 1.473223 & 1.897829 & 1.533934 & 1.287135 & 2.240486 & 1.799508 & 1.503571 & 4 & 2 \\
        64 & 2.332749 & 1.806008 & 1.473311 & 1.897833 & 1.533936 & 1.287137 & 2.240469 & 1.799498 & 1.503564 & 4 & 2 \\
        128 & 2.332955 & 1.806137 & 1.473397 & 1.897827 & 1.533933 & 1.287135 & 2.240469 & 1.799497 & 1.503564 & 4 & 2 \\
        \bottomrule
    \end{tabularx}
    \caption{Compression rates, $\text{CR}_c$, for selected compression parameters, ordered by the state size.}
    \label{table:compression_rate}
\end{table*}

\begin{table*}[htb]
    \centering
    \begin{tabularx}{\linewidth}{ccccccccccc}
        \toprule
        \bf{State Size [MB]} & \bf{Uncompressed} & \bf{FPZIP 32} & \bf{FPZIP 40} & \bf{FPZIP 48} & \bf{ZFP 32} & \bf{ZFP 40} & \bf{ZFP 48} & \bf{ZFP 1e-6} & \bf{ZFP 1e-8} & \bf{ZFP 1e-10} \\
        \midrule
        \multicolumn{11}{c}{\bf{Load State from PFS (median) [ms]}} \\
        \midrule
        16 & 877.1 & 776.8 & 769.6 & 779.0 & 805.3 & 820.3 & 824.3 & - & - & - \\
        32 & 933.8 & 793.7 & 848.2 & 860.3 & 872.2 & 907.5 & 926.7 & 810.1 & 854.6 & 871.3 \\
        64 & 989.6 & 957.5 & 1011.9 & 1052.4 & 845.0 & 892.4 & 897.7 & 877.6 & 945.7 & 948.2 \\
        128 & 1185.0 & 954.5 & 1030.5 & 1119.2 & 957.9 & 995.7 & 1011.9 & 977.3 & 1045.6 & 1044.4 \\
        \midrule
        \multicolumn{11}{c}{\bf{Speedup Load [\%]}} \\
        \midrule
        16  & - & 11.4 & 12.3 & 11.2 & 8.2 & 6.5 & 6.0 & - & - & - \\
        32  & - & 15.0 & 9.2 & 7.9 & 6.6 & 2.8 & 0.8 & 13.2 & 8.5 & 6.7 \\
        64  & - & 3.2 & 2.3 & 6.3 & 14.6 & 9.8 & 9.3 & 11.3 & 4.4 & 4.2 \\
        128 & - & 19.4 & 13.0 & 5.6 & 19.2 & 16.0 & 14.6 & 17.5 & 11.8 & 11.9 \\
        \midrule
        \multicolumn{11}{c}{\bf{Store State to PFS (median) [ms]}} \\
        \midrule
        16 & 524.9 & 479.6 & 481.2 & 491.2 & 500.0 & 501.3 & 516.3 & - & - & - \\
        32 & 651.0 & 513.7 & 527.4 & 567.9 & 524.7 & 540.0 & 593.2 & 514.3 & 531.6 & 588.2 \\
        64 & 761.0 & 585.4 & 599.2 & 659.0 & 461.1 & 498.4 & 534.9 & 582.6 & 590.0 & 653.3 \\
        128 & 1215.5 & 744.8 & 966.7 & 1132.2 & 698.2 & 869.9 & 956.4 & 676.9 & 833.3 & 943.8 \\
        \midrule
        \multicolumn{11}{c}{\bf{Speedup Store [\%]}} \\
        \midrule
        16  & - & 8.6 & 8.3 & 6.4 & 4.7 & 4.5 & 1.6 & - & - & - \\
        32  & - & 21.1 & 19.0 & 12.8 & 19.4 & 17.1 & 8.9 & 21.0 & 18.3 & 9.6 \\
        64  & - & 23.1 & 21.3 & 13.4 & 39.4 & 34.5 & 29.7 & 23.4 & 22.5 & 14.2 \\
        128 & - & 38.7 & 20.5 & 6.9 & 42.6 & 28.4 & 21.3 & 44.3 & 31.4 & 22.4 \\
        \bottomrule
    \end{tabularx}
    \caption{Speedup for the various compression parameters while storing and loading the states fomr the PFS.}
    \label{table:load_store}
\end{table*}

\begin{table*}[b]
    \centering
    \begin{tabularx}{\linewidth}{ccccccccccc}
        \toprule
        \bf{Qualifier} & \bf{FPZIP 32} & \bf{FPZIP 40} & \bf{FPZIP 48} & \bf{ZFP 32} & \bf{ZFP 40} & \bf{ZFP 48} & \bf{ZFP 1e-6} & \bf{ZFP 1e-8} & \bf{ZFP 1e-10} & \bf{SP} \\
        \midrule
        $\rho_{X_c}$                & 0.67014 & 0.98790 & 0.99995 & 0.98595 & 0.99994 & 1 & 0.85887 & 0.99574 & 0.99998 & 0.91457 \vspace{0.2cm} \\
        $\Delta\text{RMSZ}_{X_c}^p$ & 0.22 & 0.04 & 0.0009 & 0.03 & 0.001 & 3.6e-6 & 0.21 & 0.03 & 0.0005 & 0.14 \vspace{0.2cm} \\
        $A (\text{RMSE})$       & 1.41(05)  &  1.0007(02)  &     1.0002(01)   &    0.9958(22)  & 0.9964(23) &  0.9963(23) &  1.0855(90) &  0.9965(20) &  0.9973(20) &  1.0232(29) \\
        \midrule
        \bf{OK} & \xmark & \xmark & \checkmark & \xmark & \checkmark & \checkmark & \xmark & \xmark & \checkmark & \xmark \\
        \bottomrule
    \end{tabularx}
    \caption{Summary of the best compression parameters and the exclusion criteria.}
    \label{tab:statistics_result}
\end{table*}

To evaluate our framework, we perform an exemplary workflow that a user
would follow when applying it to the climate model at stake. First, we
 use the validation mode to identify viable compression parameters.
Afterwards, we apply the parameters to performing the data
assimilation with the dynamic mode.
We instrumented the validators to acquire information about the performance of the
individual validation tasks. For measuring the performance during the
dynamic mode, we leverage the internal profiler of~\Melissa{}. We
present the analysis on the
statistical qualifiers that are computed during the validation mode
in~\Cref{subsec:statistical-evaluation}.
The performance of the validators is evaluated
in~\Cref{subsubsec:validator_performance} and finally, the evaluation
of the performance during the dynamic mode
in~\Cref{subsubsec:adapted_mode}.

\subsection{Experimental Setup}\label{subsec:experimental-setup}

All of our experiments were performed on Fugaku~\cite{fugaku}, 
a 488~(double-precision)~PFlops supercomputer hosted by RIKEN R-CCS in Japan.
Fugaku consists of 158,976 compute nodes that are each equipped with a Fujitsu A64FX CPU. A64FX provides 48 application CPU cores and is integrated with 32~GiB of HBM2 memory. The compute nodes are interconnected through the TofuD network.

Each group of 16 compute nodes in Fugaku shares a 1.6~TB SSD storage, and all the nodes can access the global 150~PB Lustre file system.
We utilize the SSDs in the so-called \textit{local} mode,
where each SSD is divided proportionally among compute nodes in the given group and is exposed as dedicated per-node file system.

\subsection{Methodology}\label{subsec:methodology}

We performed experiments with different state sizes, $N$, and
different ensemble sizes, $P$, to examine the scaling behavior of the
validation mode. We selected such $N$, that result in state sizes of 16, 32,
64 and 128 MB. We further set $P$ to 25, 50 and 100 particles, using 36, 72 and 132 compute nodes respectively.
All experiments are performed using the
Loren96~\cite{lorenz1996predictability} model. The model equation reads:
\begin{equation}
    \frac{dx_i}{dt} = (x_{i+1}-x_{i-2}) x_{i-1}-x_{i}+F \quad\,\quad i = 1,\dots,N
    \label{eq:lorenz}
\end{equation}
at a forcing of 6 or larger (i.e., $F\leq 6$), the model exhibits
chaotic behavior. Hence, we set the forcing to 6 in all our
experiments. The model has been initialized with small perturbations
at $t_0$. Before starting the data assimilation, we run the model for
a long enough time ($DT = 10$), so that it exhibits a chaotic
state. We
further introduced a generic perturbation to the
states at the beginning of each particle propagation. To ensure that
we can track the errors that are introduced by the compression
method only, we used the same static seed for all particles $p_c$ with
$c=0,\dots,C$, and $C$ the number of compression parameters.
Thus, at the beginning of each particle propagation, we reset the seed
to a value $sd=f(\text{pid}, \text{rid}, \text{rank}, \text{t})$,
with \emph{pid} the particle-id, \emph{rid} the runner-id,
\emph{rank} the mpi rank of the runner and \emph{t} the assimilation
cycle.

\subsection{Statistical Evaluation}\label{subsec:statistical-evaluation}

In this section, we present the results of our experiments in validation mode.
The aim of this analysis is to determine viable
compression parameters for production runs. For this, we will
evaluate the statistical qualifiers presented
in~\Cref{subsec:validate-mode}. The qualifiers have to fulfill
certain requirements which we will pose in the next paragraphs. If all
the requirements are met, the parameter can safely be applied for compression in the climate model.

\subsubsection{Z-Value Deviation}\label{subsubsec:z_value_deviation}

To test the ensemble consistency, we
apply the Z-test from Baker et al.~\cite{baker_methodology_2014}.
For this, we plot the Z-value deviation:

\begin{equation}
    \Delta\text{RMSZ}_{X_c}^p = \left|\text{RMSZ}_{X_c}^p-\text{RMSZ}_{X{_0}}^p\right|
    \label{eq:z_value_deviation}
\end{equation}

\Cref{fig:z_value} shows the plots resolved by compression method.
The plots show the values for cycles 1 to 18 of the ensemble data
assimilation, indicated by the different colors. The variation of the
z-value deviation can be read from the error bars of the boxes. Baker
et al.~\cite{baker_methodology_2014} requires the deviation to be smaller
than 0.1:
\begin{equation}
    0 \leq \Delta\text{RMSZ}_{X_c}^p < 0.1
    \label{eq:z_value_deviation_max}
\end{equation}
Thus, as long as all values stay within this interval,
the test is considered passed.

\subsubsection{Normalized Error Statistic}\label{subsubsec:normalized_error_statistics}
To get information on the error propagation, we plot the normalized
pointwise maximum error (NPME) and normalized root mean squared error (NRMSE):
\begin{align}
   \text{NRMSE}_{X_c} &= \frac{\text{RMSE}_{X_c}}
   {\max\left( x_{c,i} \right)-\min\left( x_{c,i} \right)}
   \quad \, \quad i = 1, 2, \dots, N \\
   \text{NPME}_{X_c} &= \frac{\max\left( \left| x_{c,i} - x_{0,i} \right| \right)}
   {\max\left( x_{c,i} \right)-\min\left( x_{c,i} \right)}
   \quad \, \quad i = 1, 2, \dots, N
\end{align}

\begin{figure}[htb]
    \centering
    \subfloat[FPZIP, HP, SP\label{fig:noise_nrmse_fpzip_hp_sp}]{%
        \includegraphics[width=0.95\linewidth]{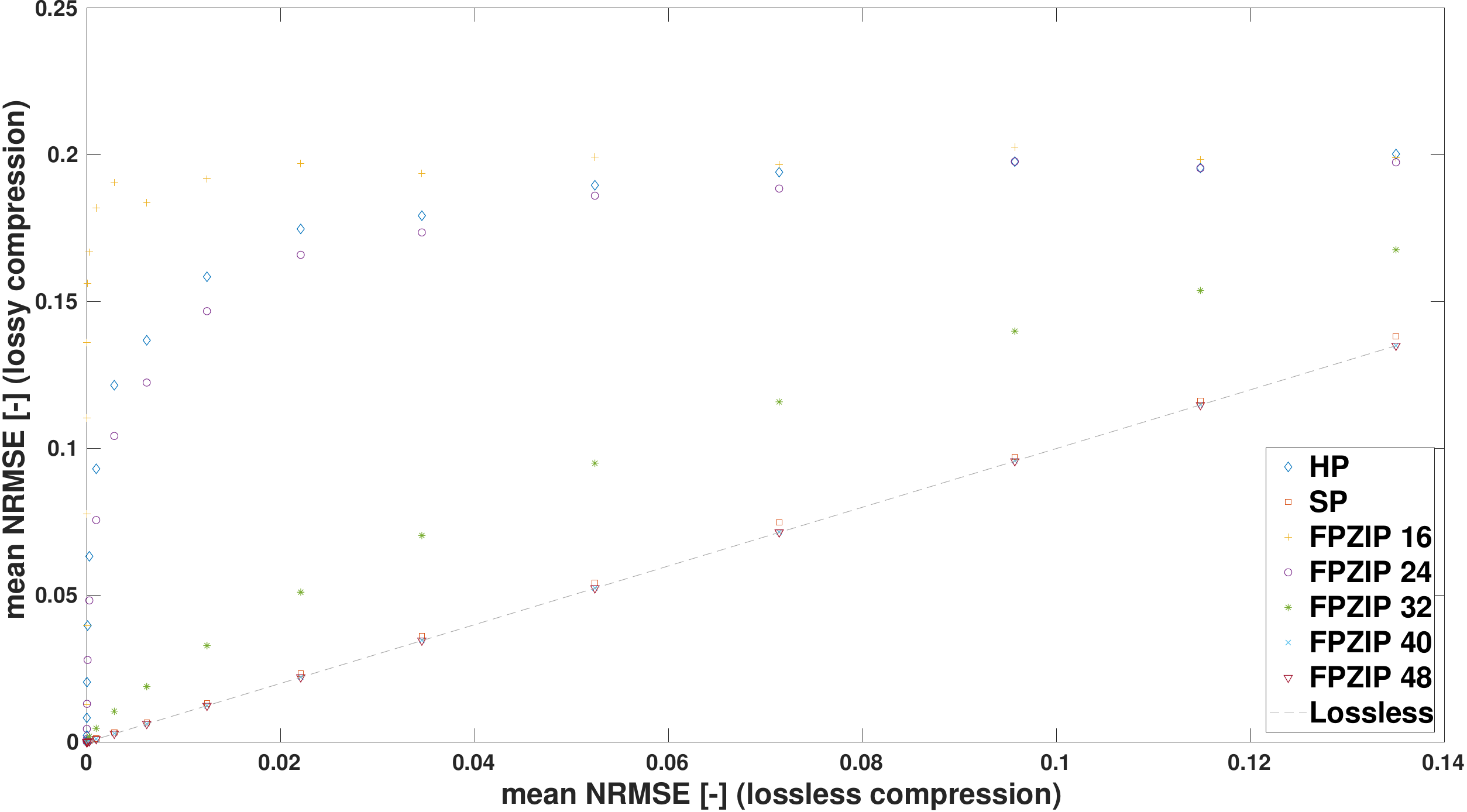}} \\
    \subfloat[ZFP\label{fig:noise_nrmse_zfp}]{%
        \includegraphics[width=0.95\linewidth]{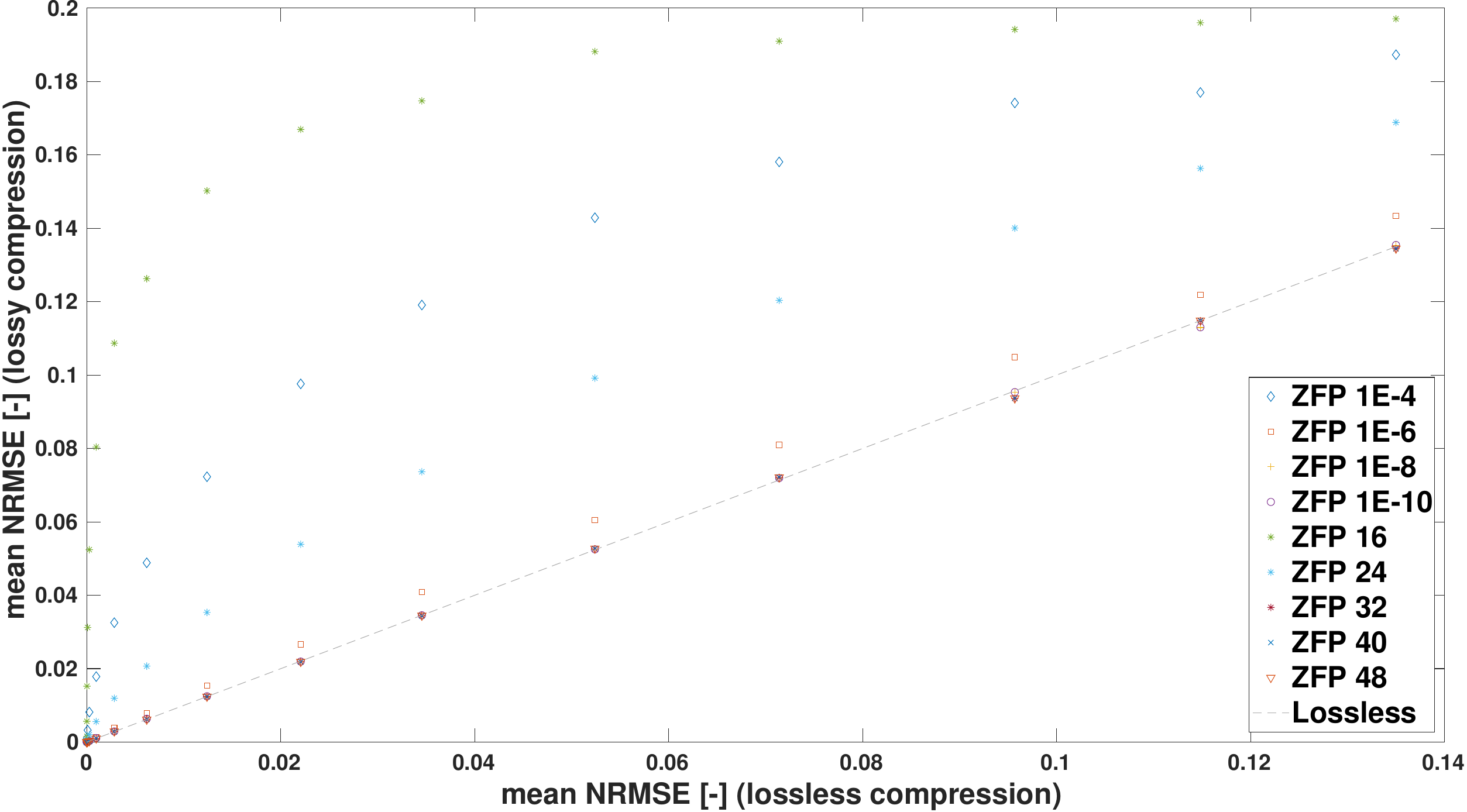}}
    \caption{Linear correlation between the timely evolution of the NRMSE of compressed to lossy compressed states. We plot the NRMSE of the compressed states for each cycle by the respective values for the lossy compressed state.}
    \label{fig:noise_nrmse}
\end{figure}

The plots are presented in~\Cref{fig:npme_nrmse_fpzip_hp_sp,fig:npme_nrmse_zfp}.
We can see that the errors increase very
quickly in the beginning and much slower towards the end. However, as
we described in~\Cref{subsec:experimental-setup}, we use the same
static seed for the propagation of identical particles. Therefore,
the errors that we see in the plots are introduced only by the
compression method. However, the randomization of the model introduces an error to
the states as well, which limits the deterioration introduced by the compression. To make the
actual impact of the compression method apparent,
we repeated the experiments for the 25 particles ensemble, adding a
small additional perturbation of O($10^{-8}$) to the states using a
random seed. Plotting the NRMSE of the various compression methods
versus that of the lossless compressed state, gives us a graphical means
to decide whether the compression method adds additional error or not.
Since the perturbation is indeed random, even the comparison between two
identical uncompressed particles differs. However, if the evolution
of the NRMSE of a compressed particle is perfectly linear to the one
of the lossless compressed states, we
can infer that no \emph{additional} error is imposed by the compression method.

\Cref{fig:noise_nrmse} shows the correlations for all compression
methods. We can see that ZFP 32, ZFP 40, ZFP 48, ZFP 1e-8, ZFP 1e-10,
FPZIP 40, FPZIP 48 and SP show an almost perfect linear relationship
to the uncompressed state. To quantify the correlation and to develop
an exclusion criteria, we performed a linear regression of the NRMSE-evolution
for all $P(P-1)/2$ combinations of the 25 particles (i.e., $P=25$), which have been lossless compressed with
FPZIP. In that way, we can determine the variation of
the linear correlation between identical states. This results into an
average correlation of $A = 1.00(01)$, with $A$ being the slope of the linear regression model $y \sim x$. Thus, we consider the NRMSE
test passed, when the value for the linear correlation lies within
the error interval of A.

\subsubsection{Pearson Correlation Coefficient}\label{subsubsec:pearson_correlation}

The last value that we look at is the Pearson correlation 
coefficient, \Cref{eq:pearson_correlation_coefficient},
$\rho_{X_c}$ encodes the linear correlation between the
compressed and uncompressed state. Baker et al.~\cite{baker_methodology_2014} requires this value to be at least
$0.99999$.

\subsubsection{Summary of the Validation Study}
A detailed list of the results is provided in~\Cref{table:statistics},
resolved by the compression method and assimilation cycle.
We list the compression rates in~\Cref{table:compression_rate}.
The results show that the rate does practically not change with increasing data size.
By matching the results with the requirements, we can
extract a subset of viable compression parameters from the initial
set. We take a closer look on the parameters that show most
promising in~\Cref{table:statistics}, and~\Cref{tab:statistics_result} summarizes the
results of the tests for the best parameters for cycle 15.
Parameters that pass the tests, indicated by the check-mark, can be
safely used for data compression in the Lorenz96 model within 15 assimilation cycles.

\subsection{Performance}\label{subsec:performance}

Now that we have outlined how to use the framework and have performed
an exemplary evaluation, we will present a performance profile of
the framework and look into the savings that we achieve during the dynamic mode.

\subsubsection{Validation Mode}\label{subsubsec:validator_performance}
The Validator parallelization is two fold. First, we parallelize among
the validator instances by distributing the particle IDs among all
available validators. Thus, each
validator computes the statistical qualifiers for $P/V$ particles,
where $P$ is the total number of particles in the ensemble and $V$
the number of validators. The number of validators can be set in the
\Melissa{} configuration. Each validator is executed on one node and
each instance is independent from the others (i.e., different MPI
executions). Second, we parallelize the validation tasks on the
validators, using the python multiprocessing class.
Hence, we compute the statistical quantities leveraging all
cores available on the node. Furthermore, the validators are
connected via TCP to each other, leveraging
ZMQ~\cite{noauthor_zeromq_nodate}, to gather all
the qualifiers on a single validator and collecting them in a
single file.

\begin{figure}[htb]
    \centering
    \includegraphics[width=\linewidth]{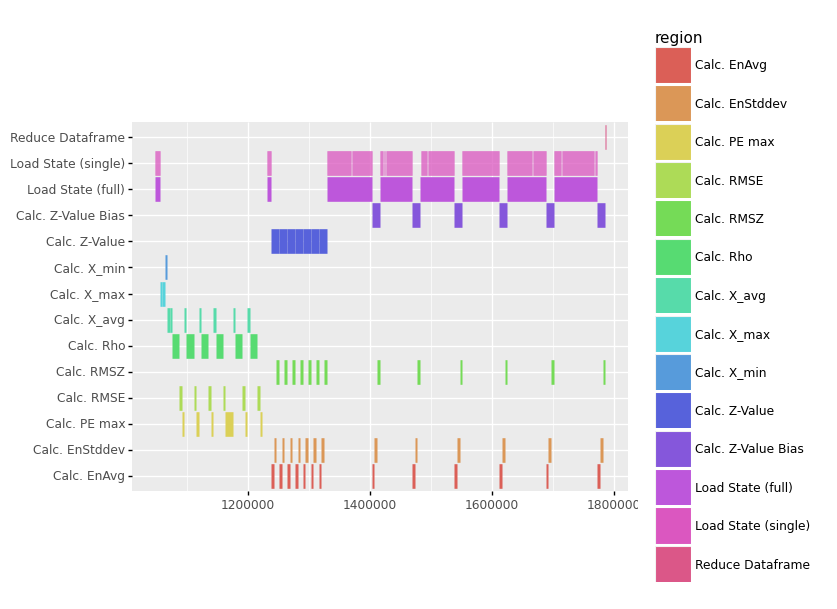}
    \caption{Trace of randomly selected validator for one validation
    cycle}
    \label{fig:trace_validator}
\end{figure}

\Cref{fig:trace_validator} shows the profile of one
validation cycle on a randomly selected validator instance for the 50
particles and 16 MB experiment.
We can see that we spend most of the time to load the particle states
(second row in the figure). The calculation of the ensemble mean and
standard deviation, require the availability of all the ensemble
particle states on the validators. Thus, we need to load in total
$(C+1)\times P$ particles on each validator. In a
first implementation, we parallelized the computations for those
quantities differently. We computed the ensemble mean and standard
deviation partially on each validator, containing only the terms for
the particles that have been assigned to the validator.
However, The terms have dimension $N$, as we can see
in~\Cref{eq:z_value}, and the reduce operation involves the transfer
of those states among the validators, even worse, we need to perform an
additional allreduce operation to compute the ensemble standard
deviation, since we need the formerly calculated ensemble mean for
this available on all validators. The
current implementation shows a considerable speedup compared to this, despite the
overhead of the additional I/O.

\begin{figure}[htb]
    \centering
    \includegraphics[width=0.95\linewidth]{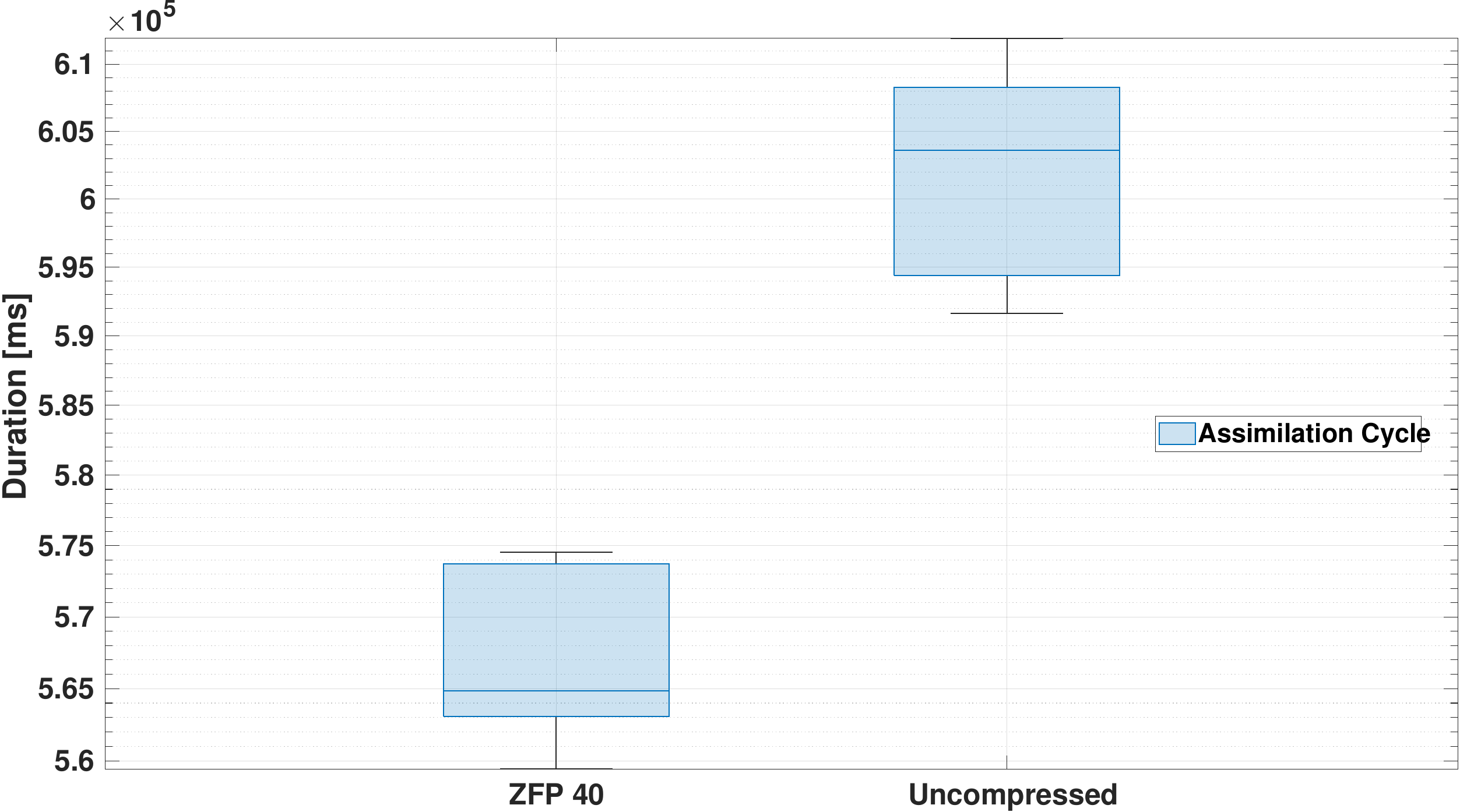}
    \caption{Comparison of the time for one assimilytion cycle leveraging the dynamic mode of our proposed framework.}
    \label{fig:dynamic_2048_iteration}
\end{figure}

\subsection{Discussion}\label{subsubsec:discussion}

\subsubsection{Dynamic mode}\label{subsubsec:adapted_mode}

In~\Cref{subsec:validate-mode}, we identified viable
compression parameters, studying the results from the validation mode.
~\Cref{table:load_store} lists the I/O performance that we measured on the runners while executing the climate model.
The times contain the compression and IO operation. We can see that we save
most while storing the states, loading the states shows a significantly smaller
speedup.
The speedup is defined by $\Delta T_{\text{SU}} = \frac{T_c-T_0}{T_0}$.
The maximum speedup is achieved with ZFP and FPZIP at 32 bit precision and with ZFP $10^{-6}$ accuracy.
The corresponding speedup is 39\%, 43\% and 44\% respectively for the
store, and 19\%, 19\% and 18\% for the load operation. The parameters
which have passed all tests also show considerable speedup for
storing and loading the states. Even the most accurate methods show a
reasonable speedup of 21\% and 22\% for storing and 15\% and 12\% for loading the states.
We further observe that the speedup increases with larger state sizes.
Moreover, we can see in~\Cref{table:compression_rate}, that the compression rate does not
change with an increasing state size. This means that we can expect even
better speedups and equal reduction rates for larger state sizes.
This is supported by experiments that we performed in dynamic mode
with a state size of 2 GB. We measured a speedup of 19\% for loading
and 57\% for storing the state. Furthermore, we have been using the
validation function, computing the maximum pointwise error of the
compressed state, while still reaching an effective speedup of 6\% for the
assimilation cycles. This can be seen
in~\Cref{fig:dynamic_2048_iteration}. The values for the speedups are
always given as the median.

Our analysis is performed on the Lorenz96 model, which has only one state
variable. Baker et al.~\cite{baker_methodology_2014} evaluates lossy
compression for four different variables of the CESM. The evaluation
shows different behavior for all variables in compression rate and
variable consistency. For instance, compression with FPZIP-24 leads
to compression rates between 2.56 and 5.26 and NRMSE values
between 1.8e-5 and 6.5e-7. Further, the ensemble consistency varies
significantly among the variables for different compression methods.
This demonstrates that we need to allow for
different compression parameters per variable and that each variable
achieves a different compression rate. That is to say, our evaluation
does not give a general statement for the performance of the
compression parameters. Moreover, it underlines the importance of
studying the impact of lossy compression on consistency for each
model and its variables separately.

\hide{
\begin{figure}[htb]
    \centering
    \includegraphics[width=\linewidth]{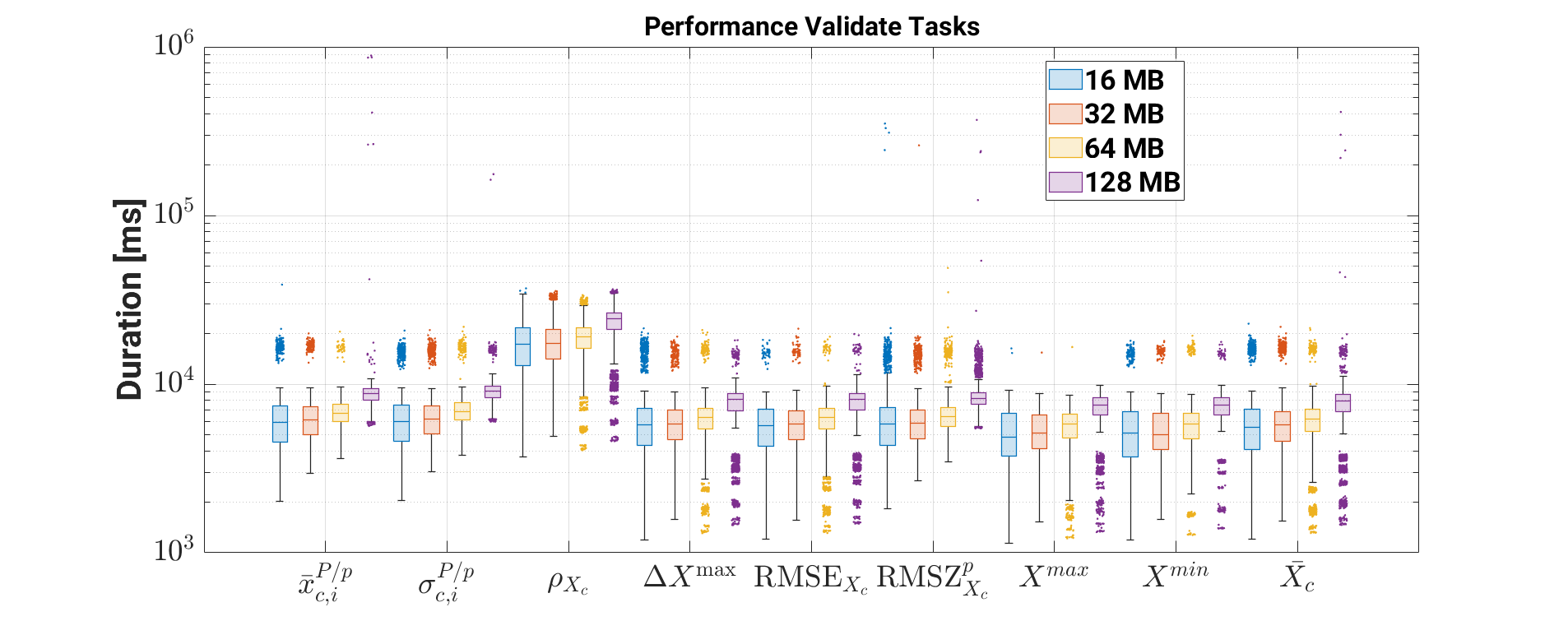}
    \caption{Performance on validator}
    \label{fig:validator_performance}
\end{figure}

\begin{table*}[htb]
    \centering
    \begin{tabularx}{\linewidth}{p{2.2cm}XXXXXXXXX}
        \toprule
        \multicolumn{10}{c}{\bf{Duration [s] (Median)}} \\
        \midrule
        \bf{State Size [MB]} & $\bar{x}_{c,i}^{P/p}$ & $\sigma_{c,i}^{P/p}$ & $\rho_{X_c}$ & $\Delta X^{\max}$ & $\textnormal{RMSE}_{X_c}$ & $\textnormal{RMSZ}_{X_c}^p$ & $X^{max}$ & $X^{min}$ & $\bar{X_c}$ \\
        \midrule
        16 & 5.93 & 5.97 & 17.29 & 5.73 & 5.68 & 5.78 & 4.85 & 5.11 & 5.52 \\
        32 & 6.13 & 6.21 & 17.41 & 5.81 & 5.81 & 5.87 & 5.12 & 5.02 & 5.71 \\
        64 & 6.71 & 6.90 & 19.01 & 6.33 & 6.32 & 6.39 & 5.80 & 5.83 & 6.17 \\
        128 & 8.77 & 9.06 & 24.39 & 8.09 & 8.12 & 8.23 & 7.52 & 7.53 & 7.98 \\
        \bottomrule
    \end{tabularx}
    \caption{MyTableCaption}
    \label{tab:validator_performance}
\end{table*}
} 

\hide{
    \begin{figure*}[htb]
        \centering
        \subfloat[FPZIP, HP, SP\label{fig:load_states}]{%
            \includegraphics[width=0.5\textwidth]{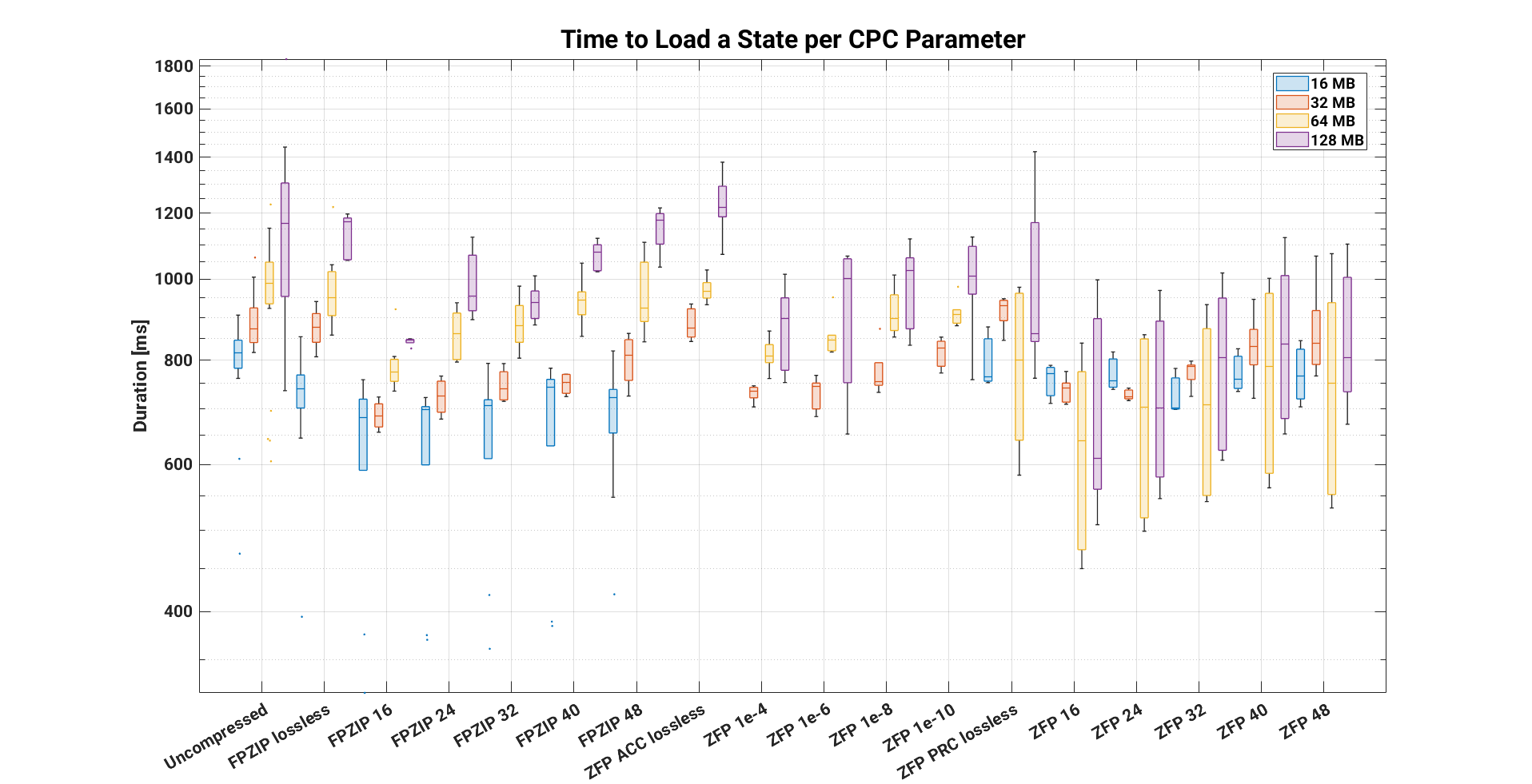}}
        \subfloat[ZFP\label{fig:store_states}]{%
            \includegraphics[width=0.5\textwidth]{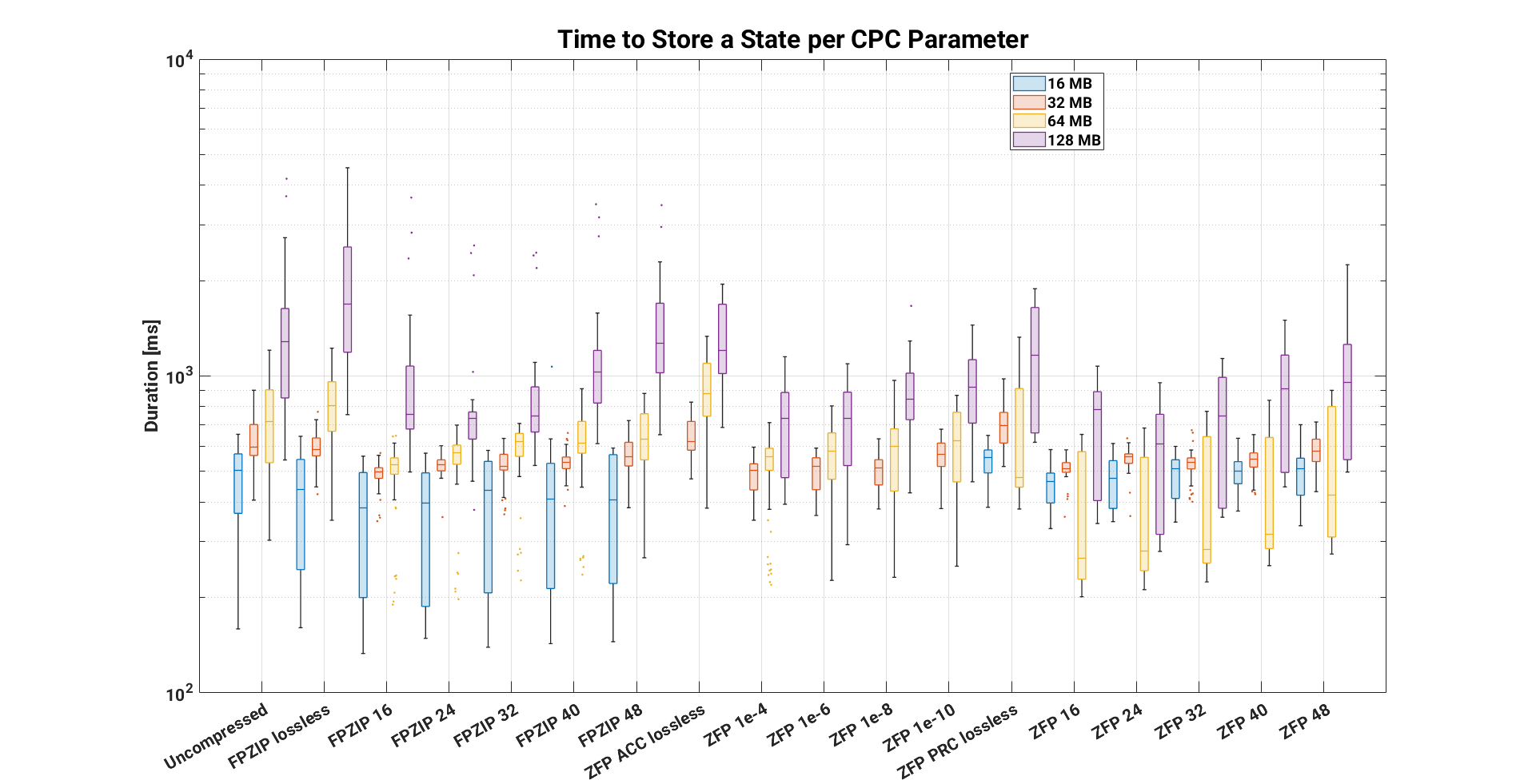}}
        \caption{Z-Value deviation,
            $\Delta\text{RMSZ}_{X_c}^p$ (see ~\Cref{eq:z_value_deviation})
            , by
            compression method. The colors indicate the different values at
            cycles from 1 to 18.}
        \label{fig:load_store_states}
    \end{figure*}
}

\hide{
    \begin{figure*}[htb]
        \centering
        \subfloat[FPZIP, HP, SP\label{fig:rho_fpzip_hp_sp}]{%
            \includegraphics[width=\textwidth]{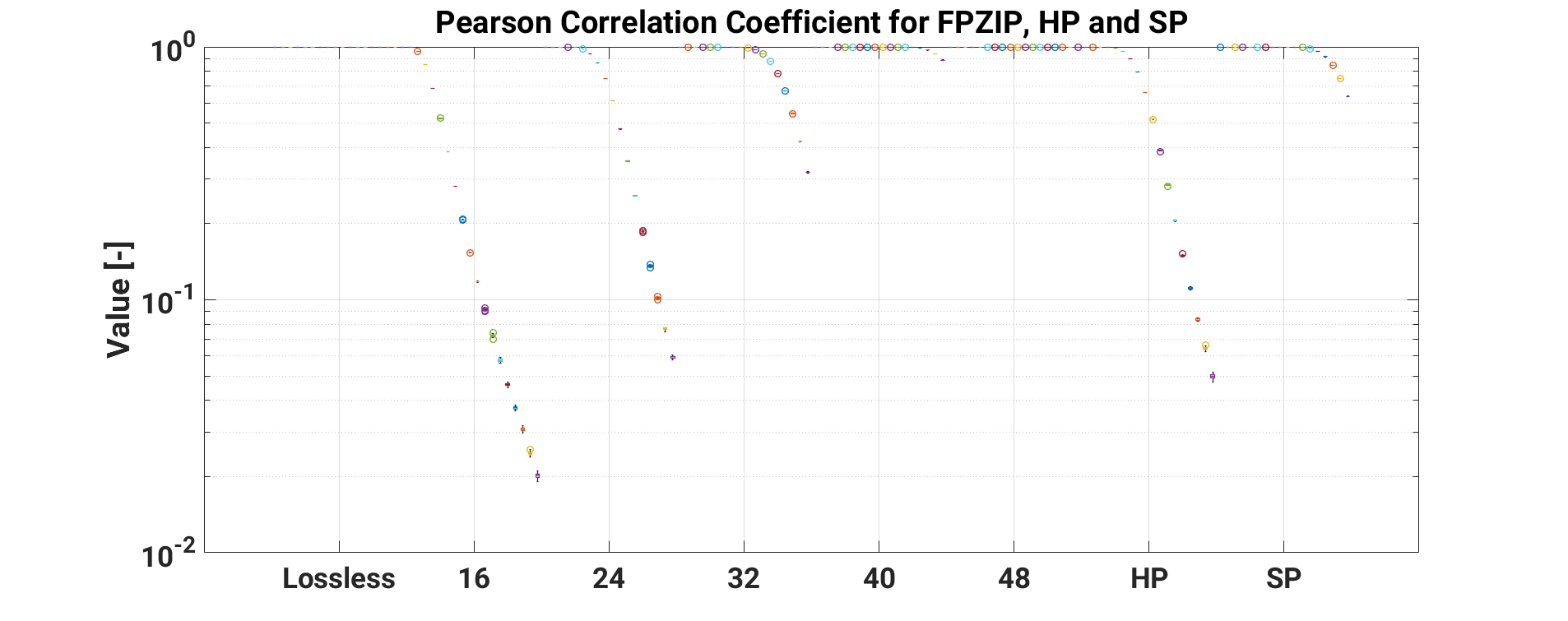}} \\
        \subfloat[ZFP\label{fig:rho_zfp}]{%
            \includegraphics[width=\textwidth]{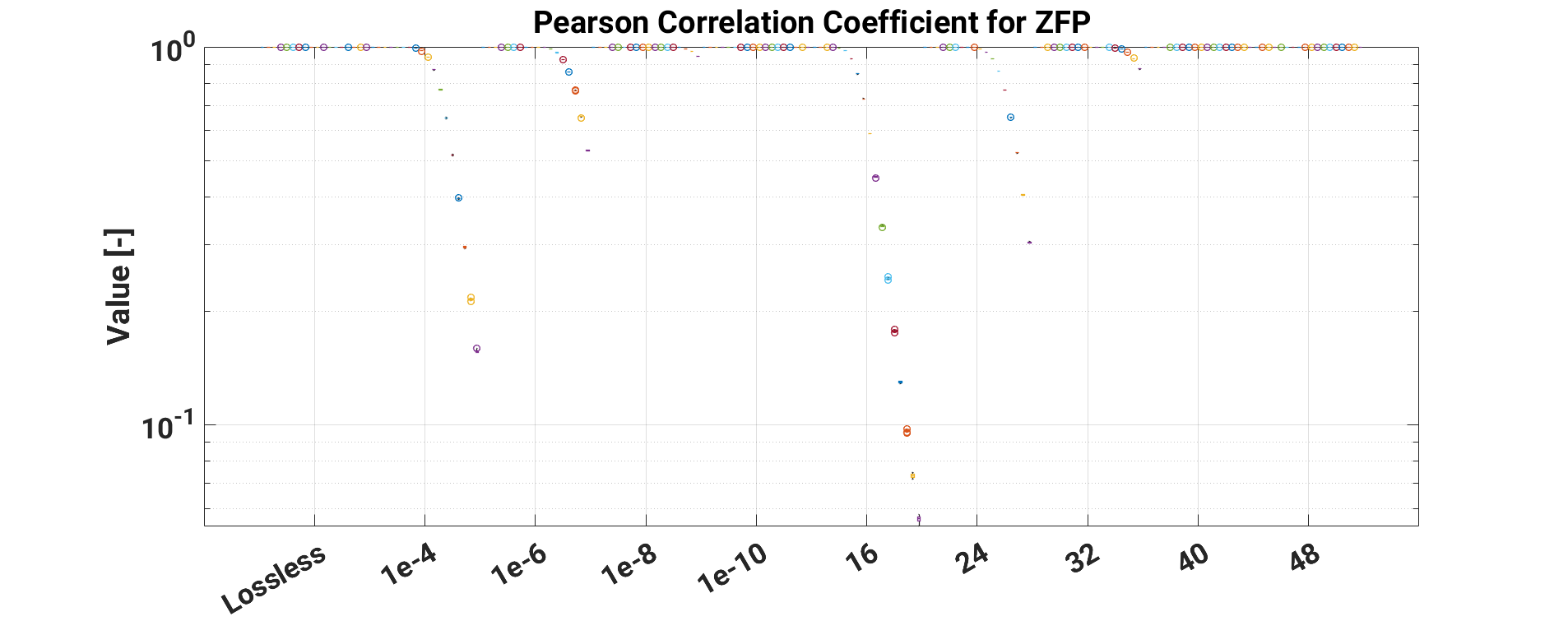}}
        \caption{Z-Value deviation,
            $\Delta\text{RMSZ}_{X_c}^p$ (see ~\Cref{eq:z_value_deviation})
            , by
            compression method. The colors indicate the different values at
            cycles from 1 to 18.}
        \label{fig:rho}

    \end{figure*}
}

\section{Related Work}\label{sec:relatedwork}

Compression of scientific datasets is not only interesting for numerical climate science. Every field in HPC that deals with large
datasets benefits from reduced data sizes. Data compression can be
applied, for instance, to datasets before visualization and to generate
checkpoints. A variety of compression algorithms are used in HPC:
ZFP~\cite{lindstrom_fixed-rate_2014}, FPZIP~\cite{lindstrom_fpzip_2017},
ISABELA~\cite{lakshminarasimhan_compressing_2011},
SZ~\cite{di_fast_2016},
MGARD~\cite{ainsworth2018multilevel} and
MGARD+~\cite{liang_mgard_2020}, to name a few. I/O libraies such as
ADIOS~\cite{godoy_adios_2020}, HDF5~\cite{noauthor_hdf5_nodate}, and
NetCDF4~\cite{delaunay_evaluation_2019} offer
high-level interfaces for compression of datasets in self-descriptive hierarchical files.

The impact and applicability of data compression to scientific
datasets has been studied in several
works~\cite{klower_compressing_2021,poppick_statistical_2020,
    delaunay_evaluation_2019,baker_methodology_2014,
    baker_evaluating_2016,prims_discriminating_2019}
The \emph{community earth system model} (CESM)
~\cite{hurrell_community_2013,danabasoglu_community_2020} includes
a Port-Validation tool, originally used to determine the
consistency of the results after porting to a different architecture.
According to Baker et al.~\cite{baker_methodology_2014}, the tool can
also be used to validate data compression for the CESM
module states. Z-Checker~\cite{tao_z-checker_2019} is a
framework that can be used to analyze the impact of compression to
any scientific dataset. The framework offers offline and online
analysis of the datasets. The online mode can be used after
instrumenting the code with the Z-Checker API functions. The online
mode can be used inside the application to observe the dynamic
behavior of data compression.

Our proposed framework is similar to the port validation tool in the
CESM, as the tool checks ensemble consistency and can detect issues
in the climate model after porting it to a new machine. However, our
framework
\begin{inparaenum}[(1)]
    \item is not constrainted to a certain climate modelling system,
    \item is more flexible as it enables definition of custom validation functions and
    \item provides automatic and direct comparison between ensembles that use different compression methods.
\end{inparaenum}
Z-Checker provides several features to evaluate the impact of the
compression method on the data, and the online mode can
potentially be used to perform an analysis that is similar to ours.
However, this would be associated with considerable implementation
efforts, and the tool does not provide measures to detect ensemble
inconsistencies.

\section{Conclusion}\label{sec:conclusion}

In this work, we present a novel framework, build on top of the
\Melissa{} architecture, that provides validation and application of
lossy compression in climate models for ensemble data assimilation.
We conducted and presented an exemplary study based on the Loren96
model, where we evaluated the applicability of the FPZIP and ZFP 16,
24, 32, 40 and 48 bit precision modes, the ZFP 1e-4, 1e-6, 1e-8,
1e-10 accuracy modes, and single and half precision floating point
representations for data compression. Our validation follows the
suggestion of Baker et al.~\cite{baker_methodology_2014}, requiring
the deviation of the Z-Values of compressed and uncompressed states to be less than 0.1
(\Cref{eq:z_value_deviation_max}), the pearson correlation
coefficient (\Cref{eq:pearson_correlation_coefficient}) of the
compressed and the uncompressed state to be at least 0.9999
(Baker et al. is more restrictive, requiring at least 0.99999), and the
impact on the normalized root mean squared error of compressed and
uncompressed state to be negligible
(compare~\Cref{subsubsec:normalized_error_statistics}). After matching
our results with this metric, we remain with FPZIP 48, ZFP 40 and 48
and ZFP 1e-10. According to this, those parameters can safely be
applied for the compression of the states for 15 assimilation cycles,
without affecting the data assimilation result. The compression rates
for those parameters are 1.47, 1.53 and 1.5 respectively, which
translates to a saving of 1/3 in storage space. Our measurements
during the dynamic mode and a state size of 2GB, show speedups of
19\% while loading and 57\% while storing the states. We also check
the integrity of the states applying the default validation function.
These results into an effective speedup of 6\% for the full
assimilation cycle.

\iftoggle{releaseStuffAfterDoubleBlind}{
\section{Acknowledgements}

Part of the research presented here has received funding from the
Horizon 2020 (H2020) funding framework under grant/award number: 824158;
Energy oriented Centre of Excellence II (EoCoE-II). The present
publication reflects only the authors’ views. The European Commission is
not liable for any use that might be made of the information contained
therein.
}

\newpage

\bibliographystyle{plain}
\bibliography{references}

\end{document}